\RequirePackage{fix-cm}
\documentclass[journal=iecred,manuscript=article]{achemso}
\SectionNumbersOn

\usepackage[fontsize=10pt]{fontsize}
\usepackage[version=3]{mhchem}
\usepackage[T1]{fontenc}
\usepackage[]{subfigure}
\usepackage{threeparttable}
\usepackage{bm}
\usepackage{amssymb}
\usepackage{derivative}
\usepackage[short,nocomma]{optidef}
\usepackage{amsmath}
\usepackage{multirow}
\usepackage{setspace}
\usepackage{xcolor}

\newcommand{\tran}{{\mkern-1.5mu\mathsf{T}}}
\newcommand{\diff}{\mathop{}\!\mathrm{d}}
\newcounter{remark}

\author{Sebastián Espinel-Ríos}
\email{s.espinelrios@princeton.edu}
\affiliation{Department of Chemical
and Biological Engineering, Princeton University, Princeton, New Jersey, 08544, United States}
\author{José L. Avalos}
\affiliation{Department of Chemical and Biological Engineering, Princeton University, Princeton, New Jersey, 08544, United States}
\alsoaffiliation{Omenn-Darling Bioengineering Institute, Princeton University, Princeton, New Jersey, 08544, United States}
\alsoaffiliation
{The Andlinger Center for Energy and the Environment, Princeton University, Princeton, New Jersey, 08544, United States}
\alsoaffiliation
{High Meadows Environmental Institute, Princeton University, Princeton, New Jersey, 08544, United States}

\title[]{Hybrid physics-informed metabolic cybergenetics: process rates augmented with machine-learning surrogates informed by flux balance analysis}

\keywords{cybergenetics, metabolic engineering, machine learning, predictive control, estimation, optogenetics}

\begin{document}
\setstretch{1}

\begin{tocentry}
    \includegraphics[]{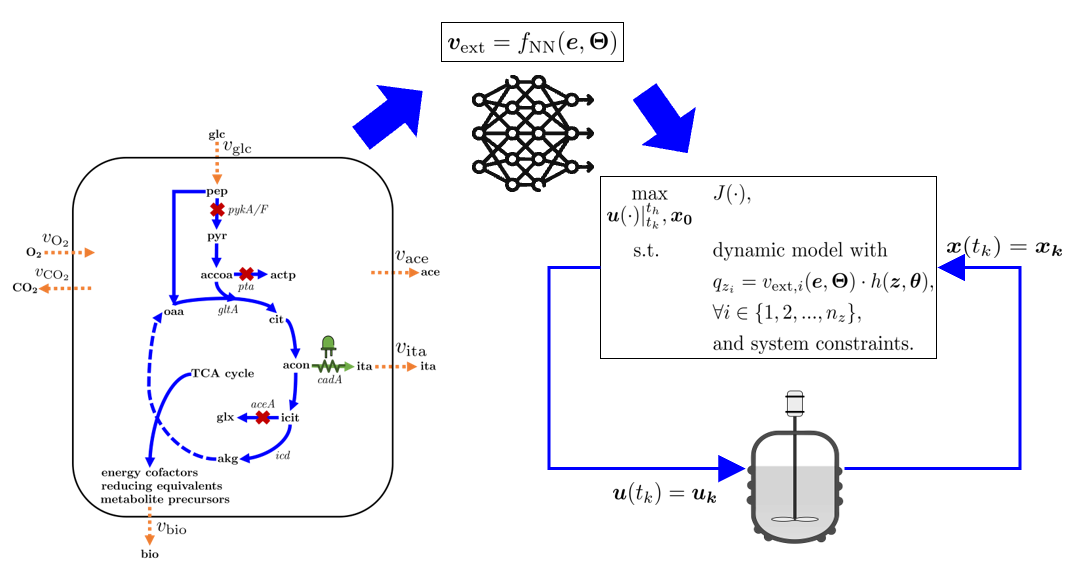}
\end{tocentry}

\begin{abstract}
\noindent
Metabolic cybergenetics is a promising concept that interfaces gene expression and cellular metabolism with computers for real-time dynamic metabolic control. The focus is on control at the transcriptional level, serving as a means to modulate intracellular metabolic fluxes. Recent strategies in this field have employed constraint-based dynamic models for process optimization, control, and estimation. However, this results in \textit{bilevel} dynamic optimization problems, which pose considerable numerical and conceptual challenges. In this study, we present an alternative hybrid physics-informed dynamic modeling framework for metabolic cybergenetics, aimed at simplifying optimization, control, and estimation tasks. By utilizing machine-learning surrogates, our approach effectively embeds the \textit{physics} of metabolic networks into the process rates of structurally simpler macro-kinetic models coupled with gene expression. These surrogates, informed by flux balance analysis, link the domains of manipulatable intracellular enzymes to metabolic exchange fluxes. This ensures that critical knowledge captured by the system's metabolic network is preserved. The resulting models can be integrated into metabolic cybergenetic schemes involving \textit{single-level} optimizations. Additionally, the hybrid modeling approach maintains the number of system states at a necessary minimum, easing the burden of process monitoring and estimation. Our hybrid physics-informed metabolic cybergenetic framework is demonstrated using a computational case study on the optogenetically-assisted production of itaconate by \textit{Escherichia coli}.
\end{abstract}

\clearpage
\section{Introduction}
Microbial biotechnology seeks to exploit the metabolic \textit{machinery} of cells through genetic and metabolic engineering to produce valuable products from renewable resources \cite{cho_designing_2022}. In this context, dynamic metabolic engineering strategies can be implemented to enhance the production efficiency of (heterologous) metabolites and proteins \cite{hartline_dynamic_2021}. Compared to \textit{static} metabolic engineering, \textit{dynamic} metabolic engineering aims to adjust the cell's metabolic flux distribution over time, offering more optimization degrees of freedom and flexibility for effectively addressing intrinsic metabolic trade-offs in bioprocesses.

A typical metabolic trade-off occurs when an increase in product yield diverts resources from biomass synthesis, consequently reducing biomass yields and volumetric productivity rates \cite{banerjee_perspectives_2023}. To address this issue, dynamic transitions between metabolic states, such as those from growth to production, can lead to improved volumetric productivity in bioprocesses \cite{lalwani_current_2018, hartline_dynamic_2021, banerjee_perspectives_2023}. Dynamic metabolic engineering can also be considered for modulating the flux through production pathways toward achieving a desired product composition.

There are different strategies to enable dynamic metabolic engineering \cite{lalwani_current_2018, hartline_dynamic_2021, banerjee_perspectives_2023}. At the transcriptional level, the most widely implemented strategy, the gene transcription of key metabolic enzymes is modulated over time. With this, the intracellular concentration of these enzymes is adjusted dynamically, influencing the achievable metabolic fluxes in the cell. In this study, such key enzymes are referred to as \textit{manipulatable enzymes}, and the metabolic fluxes they catalyze as \textit{manipulatable fluxes}. At the post-translational level, metabolic modulation targets enzymes after protein synthesis, often constitutively expressed. For instance, modifying the physical clustering of constitutively expressed enzymes can impact metabolic fluxes \cite{walls_modular_2023}. However, constitutive expression of enzymes can impose a metabolic burden in cells \cite{wu_metabolic_2016}.

Transcriptional gene modulation will be the focus of this paper. External input signals used to modulate gene transcription range from chemicals such as IPTG \cite{xu_iptginduced_2023} and methanol \cite{vogl_regulation_2013}, to process conditions, including temperature \cite{harder_temperaturedependent_2018}, pH \cite{hoynesoconnor_enabling_2017}, and oxygen supply \cite{wichmann_characterizing_2023}. Many of these inducers, nevertheless, are irreversible, i.e., it is challenging to remove them or reduce their levels once triggered. Their effective delivery can also be hindered by mass- or heat-transfer limitation in bioreactors. In addition, chemical inducers might also induce cellular stress and toxicity \cite{dvorak_exacerbation_2015,lin_oxidative_2021}. Moreover, there may be unintended cellular effects due to the non-orthogonality of the inducers. For example, pH, temperature, and oxygen availability may have overreaching effects on the overall physiological and metabolic state of cells \cite{maurer_ph_2005,gadgil_transcriptional_2008,bettenbrock_towards_2014}. These issues have motivated the rise of optogenetics in dynamic metabolic engineering, employing light-responsive transcription factors to regulate gene expression \cite{carrasco-lopez_optogenetics_2020,hoffman_optogenetics_2022,wegner_bright_2022}. Light is particularly appealing as a control input signal due to its high tunability, reversibility, orthogonality, and cost-effectiveness.

\begin{figure*}[htb!]
\centering
\includegraphics[scale=0.5]{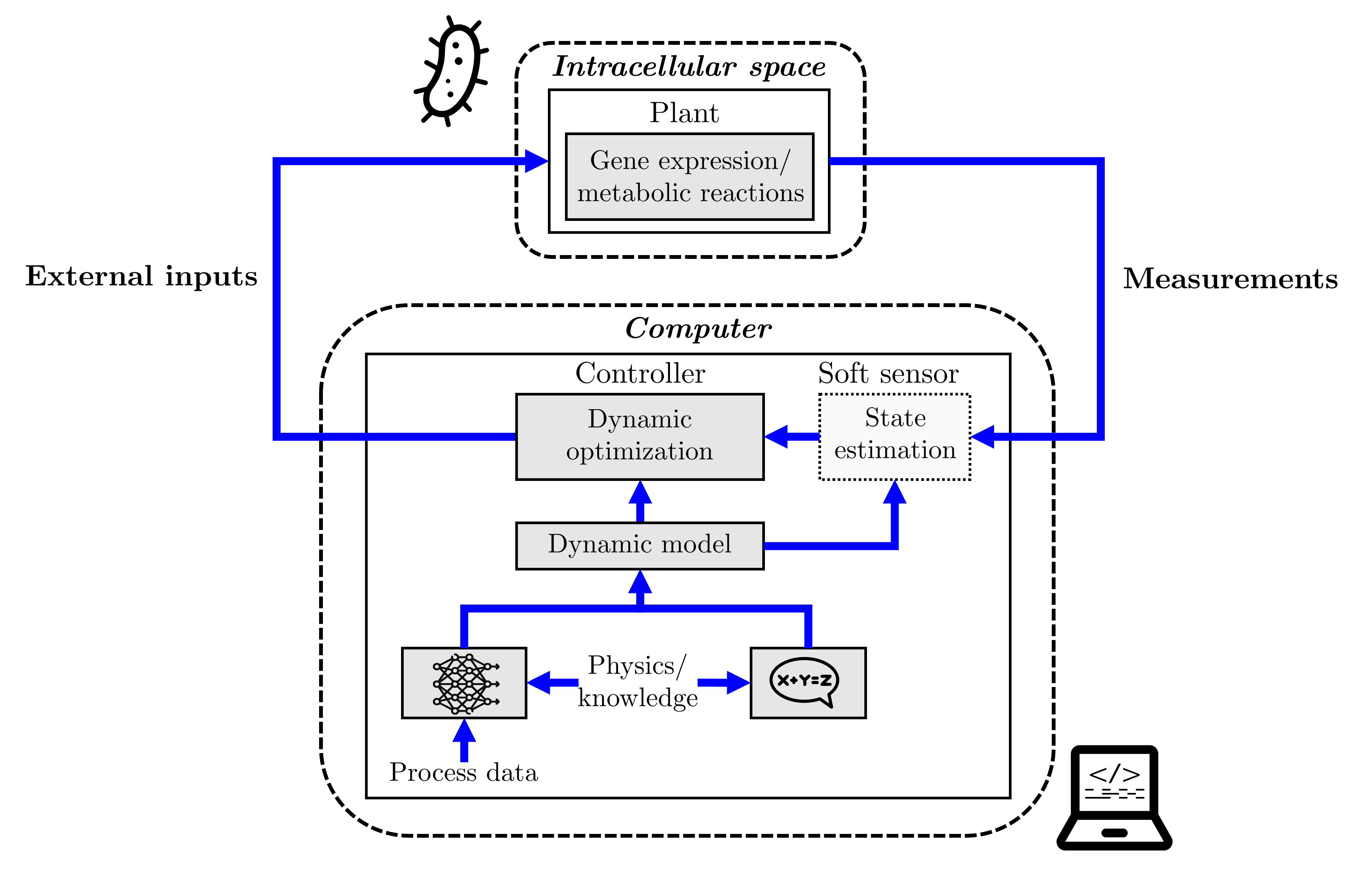}
\caption{Scheme of a metabolic cybergenetic control system. Optimal system inputs are obtained through computational methods such as dynamic optimization. These external input signals modulate the expression of metabolic enzymes and thereby metabolic fluxes. Closed-loop control is implemented by incorporating state feedback and re-optimizing the system. Optionally, if not all the state information can be measured, soft sensors in the form of state estimators can be employed to reconstruct the system state. The dynamic model can contain prior knowledge such as physics (e.g., metabolic networks) and phenomenological relations (e.g., kinetic functions), which can either be modeled explicitly or indirectly via machine-learning surrogates. Process data can also be used to capture certain parts of the model using machine learning.}
\label{fig:met_cyberg}
\end{figure*}

Determining the optimal inputs in dynamic metabolic engineering is not trivial; for example, the optimal light input profiles in optogenetically controlled systems. Model-based dynamic optimization has been shown to be a powerful tool in predicting dynamic inputs for optimally \textit{steering} the expression of manipulatable enzymes in bioprocesses \cite{espinel_cybergenetic_2023}. The application of open-loop model-based optimization often fails to consider system uncertainties, such as disturbances or model-plant mismatch. The concept of \textit{metabolic cybergenetics} (see Fig. \ref{fig:met_cyberg}), which is the focus of this paper, has emerged as a way to address the limitations of open-loop control by integrating state feedback \cite{carrasco-lopez_optogenetics_2020,espinel_cybergenetic_2023}. These systems adopt closed-loop control, such as model predictive control, by iteratively updating and solving the open-loop optimization problem with the current system state \cite{espinel_cybergenetic_2023}. Other control configurations are also possible, e.g., by merging external control via open-loop optimization and in-cell feedback encoded by genetic circuits \cite{ohkubo_hybrid_2023}.

Implementing model-based optimization for metabolic cybergenetics requires a suitable mathematical model that captures the complex dynamics of relevant metabolic processes. The model should integrate, for example, the dynamics of the expression of manipulatable enzymes (micro-scale phenomena) and the corresponding effect on process exchange rates (macro-scale phenomena). Dynamic enzyme-cost flux balance analysis (deFBA) offers a constraint-based modeling platform for coupling the dynamics of metabolism with gene expression \cite{waldherr_dynamic_2015,jabarivelisdeh_optimization_2018, jabarivelisdeh_adaptive_2020, espinelrios_maximizing_2022, espinel_cybergenetic_2023}. These models are typically underdetermined and thus formulated as dynamic optimization problems subject to constraints on mass balances, resource allocation, and regulatory mechanisms. They assume that cells are driven to optimize a dynamic biological function, requiring for its formulation a significant understanding of the short-term and long-term cellular \textit{evolutionary} driving forces. 

Structured models such as deFBA integrate the dynamics of key extracellular species and intracellular biomass components, such as catalytic enzymes, ribosomes, and storage compounds, among others. These models give valuable insight into cellular metabolism and its underlying processes, and have been considered in previous metabolic cybergenetic frameworks \cite{espinel_cybergenetic_2023}. Nevertheless, using these models in process optimization leads to bilevel optimization problems, as the optimal control problems are constrained by the constraint-based metabolic models \cite{jabarivelisdeh_adaptive_2020,espinelrios_maximizing_2022,espinel_cybergenetic_2023}. Solving bilevel optimizations is inherently challenging both numerically and conceptually, often requiring assumptions about the objectives' collaborative interaction (e.g., \textit{optimistic} or \textit{pessimistic}), which is not trivial. In an optimistic scenario, the inner optimization, the constraint-based metabolic model, can be replaced by its Karush-Kuhn-Tucker (KKT) conditions (see Refs. \citenum{dempe_solution_2019} for more details). This substitution transforms the bilevel optimization into a single-level optimization. Unfortunately, it also introduces non-convexity due to the Lagrangian and the complementary slackness constraints of the KKT conditions. Additionally, the Lagrange multipliers from the KKT conditions must be included as additional optimization variables. Moreover, deFBA and similar models usually involve a large number of states. In feedback control schemes such as model predictive control, where one needs to update the system state at different sampling times \cite{rawlings_model_2020}, using deFBA-type models imposes significant hassles for implementation due to the large numbers of states to be monitored.

In the context of dynamic metabolic engineering, we recently introduced an alternative simpler hybrid dynamic metabolic modeling technique augmented with machine-learning surrogates \cite{espinel-rios_linking_2023}. There, we represent metabolic exchange fluxes as functions of manipulatable intracellular metabolic fluxes through neural-network surrogates informed by flux balance analysis (FBA). Classical FBA operates under steady-state assumptions while considering a \textit{static} cell's objective function \cite{benner_stoichiometric_2014}. These surrogates for metabolic exchange fluxes, which link intra- and extra-cellular metabolic domains, can be plugged into the process rates of macro-kinetic dynamic models. Using a metabolic network of \textit{Escherichia coli} as an example in our previous work \cite{espinel-rios_linking_2023}, we demonstrated that these hybrid models facilitate the determination of optimal flux profiles through \textit{single-level} dynamic optimization problems, with the manipulatable intracellular fluxes being the dynamic degrees of freedom.

A limitation of the previous hybrid metabolic modeling approach \cite{espinel-rios_linking_2023} is the assumption that manipulatable intracellular fluxes can be adjusted directly and without any delay. This simplification, however, is not applicable to transcriptional-level metabolic flux modulation, the focus of this paper. Here, we build upon our initial modeling framework by integrating the dynamics of manipulatable enzyme expression. In this study, we differentiate between gene expression in prokaryotic and eukaryotic cells to offer a more comprehensive framework. We apply enzyme-capacity relationships that map the concentration of manipulatable enzymes to their corresponding manipulatable metabolic fluxes. After systematically exploring the solution space of FBA, constrained by manipulatable enzyme concentrations, we develop neural-network surrogates with the manipulatable enzyme concentrations as input features and the FBA-predicted metabolic exchange fluxes as output labels. These surrogates informed by the considered FBA model can then be used within the process rates of macro-kinetic dynamic models augmented with manipulatable enzyme expression. Our hybrid modeling can be straightforwardly integrated into simplified single-level optimization problems for control and estimation in metabolic cybergenetics. Moreover, due to the \textit{macro-kinetic} nature of the proposed models, the number of states involved is comparatively smaller than it would be with for example deFBA-type models, simplifying process monitoring.

Previous hybrid bioprocess models aim to learn or correct process rates using \textit{process data} \cite{espinel-rios_machine_2023,rogers_investigating_2023,pinto_hybrid_2023}. Our proposed modeling strategy differs by embedding the \textit{physics} of FBA into the process rates, making it \textit{physics-informed}. However, our approach does not exclude the use of machine learning to refine uncertain parts of the model using process data, potentially in a synergistic manner. Furthermore, this work extends beyond previous polynomial-based functions for mapping sets of exchange metabolic fluxes predicted by FBA \cite{de_oliveira_nonlinear_2021}. Here we explicitly consider and exploit the intracellular domain for cybergenetic control applications in dynamic metabolic engineering.

The remainder of this paper is structured as follows. In Section \ref{sec:hyb_pi_mod}, we present our hybrid physics-informed metabolic modeling framework with manipulatable enzyme expression. Utilizing this hybrid machine-learning-supported model, we formulate in Section \ref{sec:contol_est} a feedback controller based on model predictive control aimed at maximizing the production efficiency of metabolic cybergenetic systems. We also outline an optimization-based estimator to reconstruct unmeasured intracellular states. Remark that model-based optimization, control, and estimation tasks are formulated as \textit{single-level} optimization problems. Finally, the proposed modeling and metabolic cybergenetic control framework is demonstrated with a computational case dealing with the optogenetically-assisted production of itaconate in batch fermentation by \textit{E. coli}. Open-loop optimizations are used to aid in the \textit{in silico} design of the optogenetic gene expression system. More details on the case study are given in Section \ref{sec:case_study}. 

\section{Hybrid physics-informed dynamic modeling for metabolic cybergenetics}
\label{sec:hyb_pi_mod}
We introduce our hybrid physic-informed dynamic modeling framework for metabolic cybergenetics distinguishing between prokaryotic and eukaryotic cells.

\subsection{Model structure for prokaryotic cells}
Let the external states, encompassing substrates, products, and biomass dry weight, be denoted as $\bm{z} \in \mathbb{R}^{n_z}$. The biomass dry weight is indicated with $b \in \mathbb{R}$. The manipulatable intracellular enzymes are represented by $\bm{e} \in \mathbb{R}^{n_e}$, while the external inputs for inducing gene expression are denoted by $\bm{u} \in \mathbb{R}^{n_u}$.

\begin{figure}[htb!]
    \begin{center}
        \subfigure[\textbf{Prokaryotic cell}]{\includegraphics[scale=0.40]{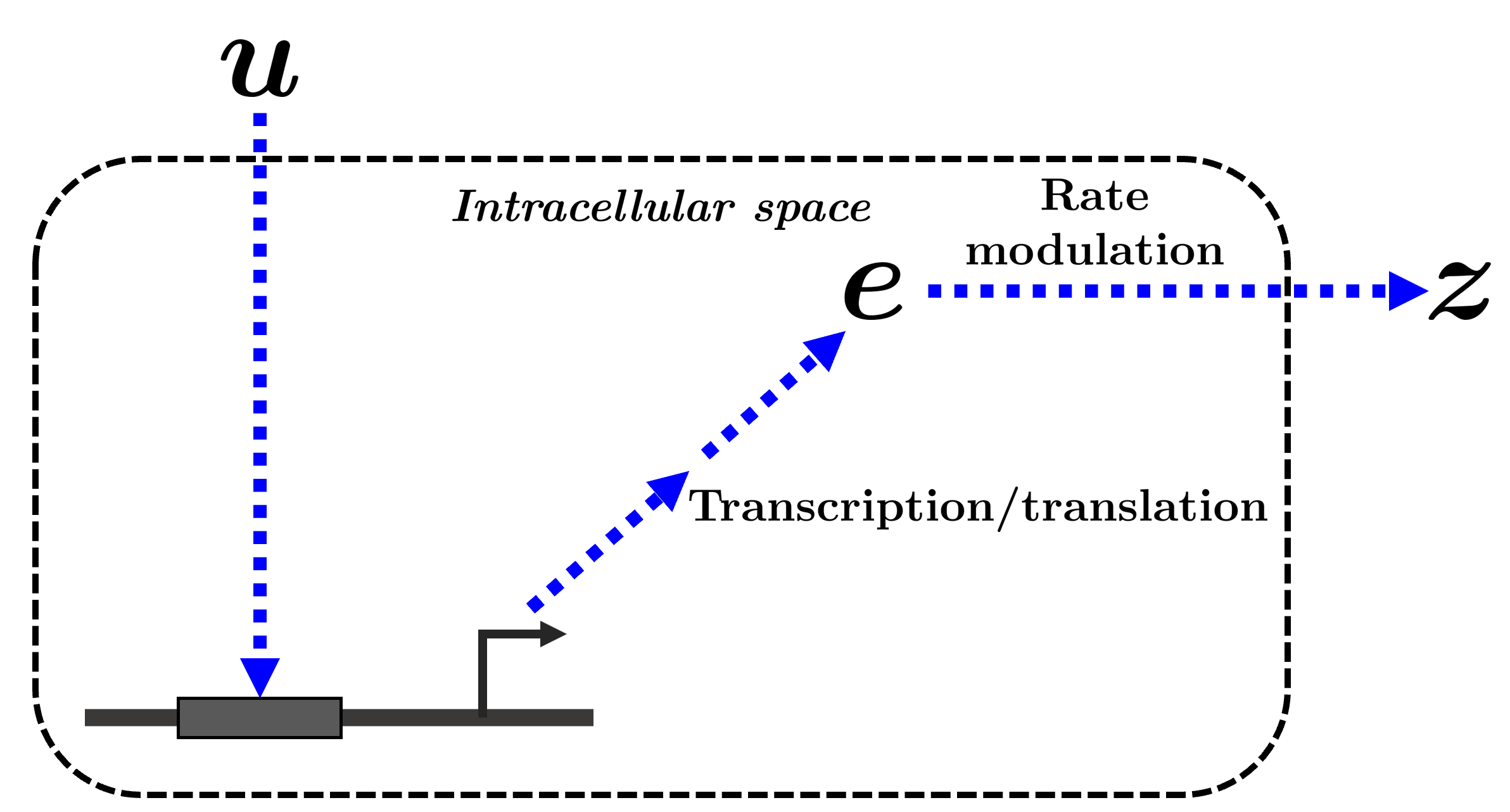}}
        \subfigure[\textbf{Eukaryotic cell}]{\includegraphics[scale=0.40]{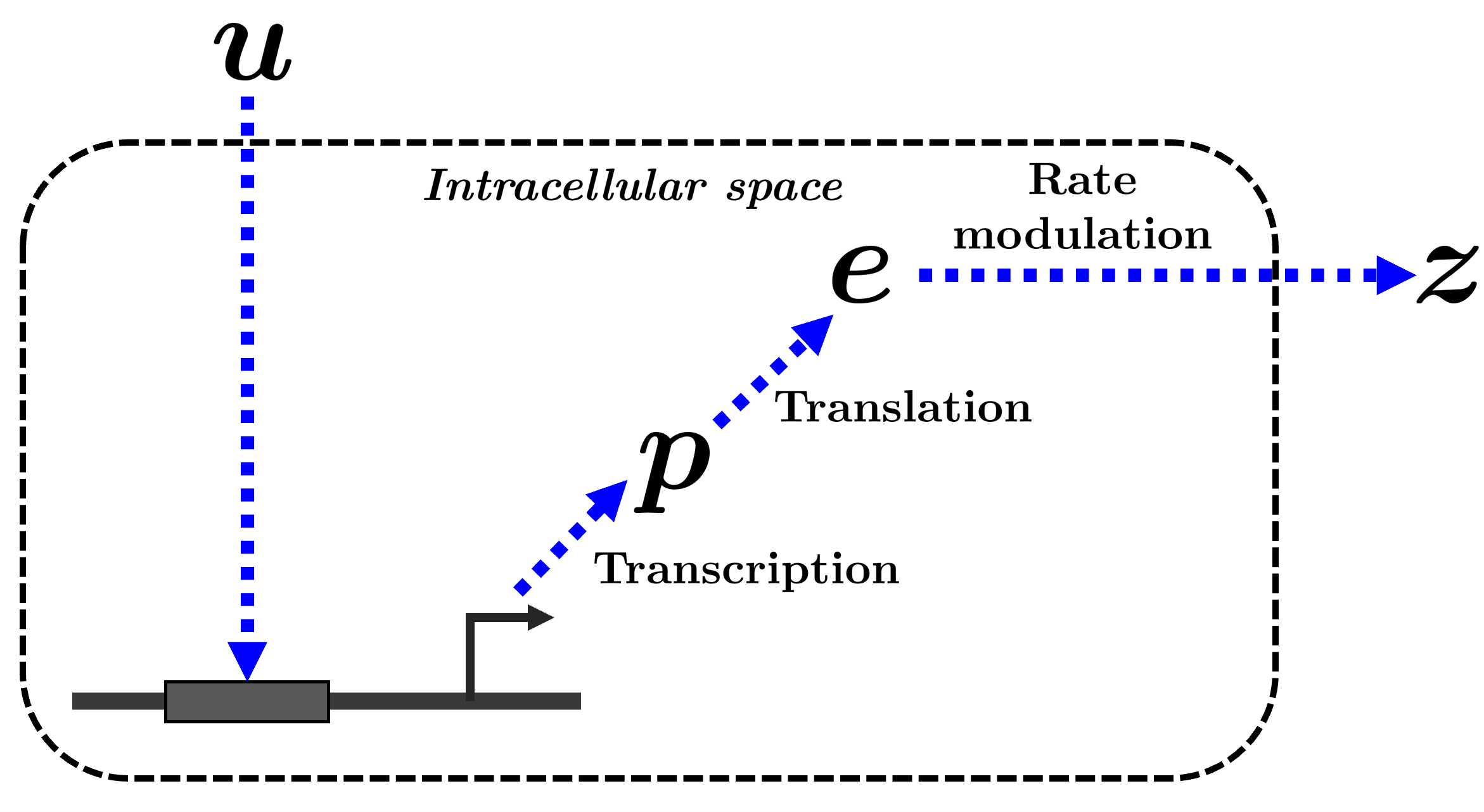}}
        \caption{Flow diagram showing the interdependence of the external inputs and the dynamic states of the proposed hybrid physics-informed cybergenetic model for (a) prokaryotic and (b) eukaryotic cells. Refer to the text for details on the notation.}\label{fig:cell_prok_euk}
    \end{center}
\end{figure}

We lump up transcription and translation processes in prokaryotic cells. In prokaryotes such as \textit{E. coli}, we deem this a valid assumption since transcription and translation occur simultaneously in the cytoplasm, i.e., they are coupled in space and time \cite{proshkin_cooperation_2010}. Therefore, for metabolic cybergenetics using prokaryotic cells, we model a batch process as follows (cf. Fig. \ref{fig:cell_prok_euk}-a):
\begin{equation}
\odv{\bm{z}(t)}{t} = b(t) \cdot \bm{q_z}(\bm{z}(t),\bm{e}(t),\bm{\theta},\bm{\Theta}), \label{eq:z_dyn} 
\end{equation}
\begin{equation}
\odv{\bm{e}(t)}{t} = \bm{q_{e_p}}(\bm{u}(t),\bm{\theta}) - \left( \mu(\bm{z}(t),\bm{e}(t),\bm{\theta},\bm{\Theta}) \cdot \bm{1_n} + \bm{d_{e}}(\bm{e}(t),\bm{\theta}) \right) \circ \bm{e}(t), \label{eq:e_dyn}
\end{equation}
\begin{equation} 
\bm{z}(t_0) = \bm{z_0}, \, \bm{e}(t_0) = \bm{e_0}. \label{eq:x0}
\end{equation}

$\bm{\theta} \in \mathbb{R}^{n_{\theta}}$ encompasses constant parameters of the explicitly-modeled prior knowledge/physics. $\bm{\Theta} \in \mathbb{R}^{n_\Theta}$ represents constant parameters of the neural-network surrogates (weights and biases) informed by FBA; these correspond to the physics-informed machine-learning component of the model. The accumulation of external states depends on biomass-specific process exchange rates $\bm{q_z}: \mathbb{R}^{n_z}  \times \mathbb{R}^{n_e} \times \mathbb{R}^{n_{\theta}} \times \mathbb{R}^{n_\Theta} \rightarrow \mathbb{R}^{n_z}$. The dynamics of intracellular manipulatable enzymes in \textit{prokaryotic} cells depend on the input-dependent production rate (lumped transcription and translation) $\bm{q_{e_p}}: \mathbb{R}^{n_u} \times \mathbb{R}^{n_{\theta}} \rightarrow \mathbb{R}^{n_e}$, cell dilution given by the growth rate $\mu: \mathbb{R}^{n_z}  \times \mathbb{R}^{n_e} \times \mathbb{R}^{n_{\theta}} \times \mathbb{R}^{n_\Theta} \rightarrow \mathbb{R}$, and the degradation rate $\bm{d_{e}}: \mathbb{R}^{n_e} \times\mathbb{R}^{n_{\theta}} \rightarrow \mathbb{R}^{n_e}$. Note that $\mu \in \bm{q_z}$, $\bm{1_n} \in \mathbb{R}^{n_{e}}$ is a vector of ones, and $\circ$ indicates the Hadamard product. The initial process time is denoted as $t_0$. We assume that each external input uniquely modulates the expression of one corresponding manipulatable enzyme, thus $n_u=n_e$.  

It is worth mentioning that the external inputs affect the rates $\bm{q_{e_p}}$, influencing enzyme expression. In contrast, the process exchange rates are affected by the manipulatable enzyme concentrations, but not directly by the external inputs. This helps to capture an input-output delay in terms of the effect of the external input on the process exchange rates. In an optogenetically controlled system, the external inputs could be, for example, the light intensity or the frequency of light pulses. 

\subsection{Model structure for eukaryotic cells}
Transcription and translation in eukaryotic cells are uncoupled in space and time. Transcription takes place in the nucleus, followed by translation in the cytoplasm \cite{rodriguez-molina_knowing_2023}. Therefore, to properly account for time delays in metabolic cybergenetics using eukaryotic cells, we need to consider the dynamics of transcription separate from translation. With this in mind, we model the batch process as follows (cf. Fig. \ref{fig:cell_prok_euk}-b):
\begin{equation} 
\odv{\bm{z}(t)}{t} = b(t) \cdot \bm{q_z}(\bm{z}(t),\bm{e}(t),\bm{\theta},\bm{\Theta}), \label{eq:z_dyn_e}
\end{equation}
\begin{equation} 
\odv{\bm{e}(t)}{t} = \bm{q_{e_e}}(\bm{p}(t),\bm{\theta}) - \left( \mu(\bm{z}(t),\bm{e}(t),\bm{\theta},\bm{\Theta}) \cdot \bm{1_n} + \bm{d_{e}}(\bm{e}(t),\bm{\theta}) \right) \circ \bm{e}(t), \label{eq:e_dyn_e}
\end{equation}
\begin{equation} 
\odv{\bm{p}(t)}{t} =  
\bm{q_{p}}(\bm{u}(t),\bm{\theta}) - \left( \mu(\bm{z}(t),\bm{e}(t),\bm{\theta},\bm{\Theta}) \cdot \bm{1_n} + \bm{d_{p}}(\bm{p}(t),\bm{\theta}) \right) \circ \bm{p}(t), \label{eq:p_dyn_e}
\end{equation}
\begin{equation} 
\bm{z}(t_0) = \bm{z_0}, \, \bm{e}(t_0) = \bm{e_0}, \, \bm{p}(t_0) = \bm{p_0}. \label{eq:x0_e}
\end{equation}

In this case, $\bm{p} \in \mathbb{R}^{n_e}$ represents the product of transcription, i.e., the mRNA transcripts. The dynamics of the latter state is affected by the rates of transcription $\bm{q_{p}}: \mathbb{R}^{n_u} \times \mathbb{R}^{n_{\theta}} \rightarrow \mathbb{R}^{n_e}$, cell dilution (growth rate), and mRNA degradation $\bm{d_{p}}: \mathbb{R}^{n_p} \times\mathbb{R}^{n_{\theta}} \rightarrow \mathbb{R}^{n_e}$. In contrast to the model for prokaryotic cells, the dynamics of enzyme production in \textit{eukaryotic} cells are affected by the translation rate $\bm{q_{e_e}}: \mathbb{R}^{n_e} \times \mathbb{R}^{n_{\theta}} \rightarrow \mathbb{R}^{n_e}$, which does not directly depend on the external input, but rather on the availability of mRNA transcripts.

Throughout this work, $\bm{z}$ and $b$ are expressed in $\mathrm{mmol/L}$ and $\mathrm{g_b/L}$, respectively. $\mathrm{g_b}$ refers to grams of cell dry weight. The intracellular states $\bm{e}$ and $\bm{p}$ are both expressed in $\mathrm{mmol/g_b}$. While the models above are written for batch processes (the focus of this work), they can be expanded to other modes of operation, e.g., fed-batch or continuous processes, by incorporating proper dilution rates. From now on, we will omit the time dependency of the variables unless unclear from the context. In addition, we consistently use bold-notation to represent vectors and matrices and non-bold notation to represent scalars and unidimensional variables and functions.

\subsection{Macro-scale process exchange rates}
Defining the rates $\bm{q_z}$ requires elucidating the relationship between manipulatable enzyme concentrations and the change in the cell's metabolic flux distribution. One could try to use phenomenological relationships, but these often tend to overlook important intracellular aspects such as intrinsic metabolic trade-offs and energy/redox balancing \cite{zhang_hybrid_2019}. Therefore, we consider a hybrid physics-informed approach where:
\begin{equation}
\begin{aligned}
q_{z_i} = v_{\mathrm{ext},{i}}(\bm{e},\bm{\Theta}) \cdot h(\bm{z},\bm{\theta}), \, \forall i \in \{1,2,...,n_z\}. \label{eq:qi_def}
\end{aligned}
\end{equation}

In the above equation, $v_{\mathrm{ext},{i}}: \mathbb{R}^{n_e} \times \mathbb{R}^{n_\Theta} \rightarrow \mathbb{R}$ represents the metabolic exchange flux for an extracellular state $i$ predicted by a neural-network surrogate of FBA. $v_{\mathrm{ext},{i}}$ can be regarded as a \textit{maximum theoretical} rate value as classical FBA typically neglects limitations due to external concentrations or conditions. The function $h: \mathbb{R}^{n_z} \times \mathbb{R}^{n_{\theta}} \rightarrow \mathbb{R}$ explicitly captures possible prior knowledge on aspects such as substrate-limitation or product-inhibition that reduce the flux $v_{\mathrm{ext},{i}}$ --not captured by the FBA model, thus also not by the corresponding surrogate--. The function $h$ can in principle also integrate, partly or fully, a machine-learning component, e.g., a \textit{correction} or \textit{missing} term learned from process data. For clarity of notation, remark that we refer to $q_{z_i}$ as \textit{process (exchange) rates} and  $v_{\mathrm{ext},{i}}$ as \textit{metabolic (exchange) fluxes} throughout this work.

We collect all the metabolic exchange fluxes in $\bm{v_\mathrm{ext}} \in \mathbb{R}^{n_{\mathrm{ext}}}$. Then, the neural-network surrogates can be formulated as:
\begin{equation}
\label{eq:ML_model_ext}
\bm{v_\mathrm{ext}} = f_\mathrm{NN}(\bm{e}, \bm{\Theta}),
\end{equation}
where $\bm{v_\mathrm{ext}} = \{ v_{\mathrm{ext},i} \,|\, i \in \{1,2,...,n_z\} \}$. The function $f_\mathrm{NN}: \mathbb{R}^{n_e} \times \mathbb{R}^{n_\Theta} \rightarrow \mathbb{R}^{n_z}$ represents the neural-network surrogate that maps the manipulatable intracellular enzyme concentrations to the metabolic exchange fluxes. 

In this work, we consider feedforward neural networks, which consist of interconnected nodes organized into $n_l$ layers. The output vector $\bm{a_l} \in \mathbb{R}^{n_{a_l}}$ of a layer $l$ in a trained feedforward neural network is represented as follows \cite{pohlodek_flexible_2022}:
\begin{equation}
\bm{a_l} = \bm{\sigma_l}(\mathbf{W}_l \cdot \bm{a_{l-1}} + \bm{b_l}), \, \forall l \in \{1,2,...,n_l\}. \label{eq:fnn_form}
\end{equation}
 The features are the inputs to the first layer, i.e., $\bm{a_0}=\bm{e}$, and the labels are the outputs of the last layer, i.e., $\bm{a_{n_l}}=\bm{v_\mathrm{ext}}$. $\bm{\sigma_l}: \mathbb{R}^{n_{a_{l-1}}} \rightarrow \mathbb{R}^{n_{a_l}}$ is the layer's activation function. $\mathbf{W}_l \in \mathbb{R}^{n_{a_l} \times n_{a_{l-1}}}$ and $\bm{b_l} \in \mathbb{R}^{n_{a_l}}$ are weight matrices and bias vectors, respectively. Given the complexity of the relationships the surrogate model is expected to learn, a variety of machine-learning methods and tools for multivariate function approximation \cite{mowbray_machine_2021}, beyond just neural networks, can in principle also be used. However, the benefit of feedforward neural networks is their feature of being universal function approximations \cite{hornik_multilayer_1989}. 

\subsection{Micro-scale rates of the model}
Regarding the rates involved in the dynamics of the expression of manipulatable enzymes, with or without lumped transcription and translation, one can use, e.g., the widely-known Hill function for either input-dependent activation or repression \cite{ang_tuning_2013}. Degradation rates can be modeled, e.g., following commonly employed first-order degradation kinetics \cite{yin_kinetic_2018}. An example of these rate functions will be shown in the itaconate computational case study in Section \ref{sec:case_study}.

\subsection{Training the neural-network surrogates for enzyme-dependent metabolic exchange fluxes}
We now outline how to build the enzyme-dependent surrogates for the metabolic exchange fluxes based on the \textit{physics} of FBA. To do so, we first link manipulatable intracellular enzymes to manipulatable intracellular fluxes $\bm{v_\mathrm{man}} \in \mathbb{R}^{n_\mathrm{man}}$ via enzyme-capacity relationships. For a manipulatable enzyme $i$:\begin{equation} \label{eq:e_k_cat}
e_i =  \left | \frac{v_{\mathrm{man},{i}}}{k_{\mathrm{cat},i}} \right |, \, \forall i \in \{1,2,...,n_e\},
\end{equation}
where $\bm{k_{\mathrm{cat}}} = \{ k_{\mathrm{cat},i} \,|\, i \in \{1,2,...,n_e\} \}$ contains the catalytic constants of the manipulatable enzymes. Eq. \eqref{eq:e_k_cat} assumes that the manipulatable enzymes operate under substrate saturation conditions, close to their maximum reaction rates. In fact, in microorganisms such as \textit{E. coli}, most enzymes typically work under substrate saturation \cite{bennett_absolute_2009}. This simplification avoids incorporating intracellular metabolite concentrations in Eq. \eqref{eq:e_k_cat}, thereby keeping the model simple. Note that $|\cdot|$ indicates the absolute value operation, which is necessary to handle possible negative metabolic fluxes in case of reversible reactions.

The next step is to populate a dataset $\mathcal{D} \in \mathbb{R}^{n_g \times (n_\mathrm{ext} + n_e)}$ with input features (manipulatable enzyme concentrations) and output labels (metabolic exchange fluxes) to train our surrogate model of FBA. Using a systematic \textit{grid search} approach, we run simulations of a given FBA model across combinations of intracellular manipulatable fluxes. The \textit{exploration space} is determined by the Cartesian product $\mathbb{G}$ of the sets  $\mathbb{V}_{\mathrm{man},1}, \mathbb{V}_{\mathrm{man},2}, \ldots, \mathbb{V}_{\mathrm{man},n}$:
\begin{equation} 
\label{eq:cartesian_product}
\mathbb{G} = \mathbb{V}_{\mathrm{man},1} \times \mathbb{V}_{\mathrm{man},2} \times \ldots \times \mathbb{V}_{\mathrm{man},n},
\end{equation}
where the set $\mathbb{V}_{\mathrm{man},n}$ contains user-defined values in the \textit{grid} for the $n$-th manipulatable flux. Note that the total number of combinations of manipulatable fluxes is indicated by $n_g = \text{size}(\mathbb{G})$ and each element of $\mathbb{G}$ is an $n_\mathrm{man}$-tuple. 

Let us refer to the $j$-th element of $\mathbb{G}$ as $\bm{g_j}$. To fill the dataset  $\mathcal{D}$, the FBA model constrained by the manipulatable enzyme concentrations is systematically solved as:
\begin{align}
\max_{\bm{v},\bm{e}} \quad & F(\bm{v},\bm{e}),\label{eq:FBA_cost_k} \\
\text{s.t.} \quad & \dot{\bm{m}}=\bm{S}\bm{v} = \bm{0}, \label{eq:SV_cons_k} \\
& \bm{v_\mathrm{min}} \leq \bm{v} \leq \bm{v_\mathrm{max}}, \label{eq:V_cons_k} \\
& \bm{v_{\mathrm{man}}} = \bm{g_j}, \,\bm{v_\mathrm{man}} \subseteq \bm{v}, \label{eq:grid_cons_k} \\
& e_i = \left | \frac{v_{\mathrm{man},{i}}}{k_{\mathrm{cat},i}} \right |, \forall i \in \{1,2,...,n_e\}, \label{eq:e_cons_k}  \\
& \forall j \in \{1,2,...,n_g\}. \notag
\end{align}
$F:\mathbb{R}^{n_v} \times \mathbb{R}^{n_e} \rightarrow \mathbb{R}$ captures the assumed static objective function that the cell maximizes. The decision variables of the optimization are the (intracellular and extracellular) metabolic fluxes of the metabolic network $\bm{v} \in \mathbb{R}^{n_v}$ and the intracellular manipulatable enzymes. Furthermore,  $\bm{m} \in \mathbb{R}^{n_m}$ represents intracellular metabolites and $\bm{S} \in \mathbb{R}^{n_m \times n_v}$ is the stoichiometric matrix of the intracellular metabolites. Eq. \eqref{eq:SV_cons_k} is equal to a zero vector $\textbf{0}$ of appropriate dimension, consistent with the steady-state assumption of FBA. Feasible lower ($\bm{v_\mathrm{min}} \in \mathbb{R}^{n_v}$) and upper bounds ($\bm{v_\mathrm{max}} \in \mathbb{R}^{n_v}$) constrain the metabolic fluxes in Eq. \eqref{eq:V_cons_k}. The different combinations of intracellular manipulatable fluxes are integrated into the FBA optimization via the constraint in Eq. \eqref{eq:grid_cons_k}. The enzyme-capacity relationships between the manipulatable enzymes and manipulatable fluxes are represented by Eq. \eqref{eq:e_cons_k}. Note that, \textit{if necessary}, the FBA model's predictive accuracy can be in principle enhanced by considering additional constraints dealing with resource allocation, thermodynamics, and regulation, among other aspects \cite{beard_energy_2002,shlomi_genomescale_2007,goelzer_cell_2011}. For example, one can consider enzyme-capacity constraints as in Eq. \eqref{eq:e_cons_k} on other (non-manipulatable) fluxes as well. For simplicity, we only consider in this work enzyme-capacity constraints on the manipulatable fluxes.

The solution of the FBA is iteratively appended to the dataset as follows:
\begin{equation}
\label{eq:dataset}
\mathcal{D} \gets \mathcal{D} \cup \{ (\bm{v_{\mathrm{ext}}}^{(j)}, \bm{e}^{(j)}) \}, \, \forall j \in \{1, 2, \ldots, n_g\},
\end{equation}
where $\bm{v_{\mathrm{ext}}}^{(j)}$ and $\bm{e}^{(j)}$ represent the solution vectors of $\bm{v_{\mathrm{ext}}}$ and $\bm{e}$. The populated dataset, containing all the solutions from the FBA model under the different combinations of intracellular manipulatable fluxes in $\mathbb{G}$, is then used to train the neural-network surrogate in Eq. \eqref{eq:ML_model_ext}.

The training of the neural network is performed by iterative adaptation of the weight matrices and bias vectors (cf. Eq. \eqref{eq:fnn_form}) so that the loss function is minimized. Here we use the mean squared error (MSE) as the loss function:
\begin{equation} 
\label{eq:mse}
\mathrm{MSE} = \frac{1}{n_g \times n_z} \sum_{i=1}^{n_g \times n_z}(\mathfrak{y}_i-\hat{\mathfrak{y}}_i)^2,
\end{equation}
where $\hat{\mathfrak{y}}_i \in \mathbb{R}$ is the predicted value of the output label by the neural network for the $i$-th data entry, and $\mathfrak{y}_i \in \mathbb{R}$ is the actual output label value for the $i$-th data entry as predicted by FBA. In total, we have $n_g \times n_z$ data entries, which corresponds to having $n_g$ FBA solutions, each with $n_z$ associated output labels.  

Considering that sufficient data can be generated \textit{in silico} through extensive FBA simulations, we expect to have adequately large datasets for the effective training of the neural-network surrogates. In addition, as suggested in our previous work \cite{espinel-rios_linking_2023}, unfeasible flux combinations during the systematic FBA exploration can be \textit{flagged}, enabling the neural-network surrogates to learn to differentiate between feasible and unfeasible flux regions during training. Consequently, model-based optimization and control strategies utilizing these models can be designed to be \textit{aware} of and \textit{avoid} unfeasible regions accordingly.

The proposed surrogate approach assumes that an appropriate metabolic network is \textit{a priori} available. Note that there are several genome-scale models at hand for different microorganisms derived from gene-enzyme-reaction associations \cite{passi_genome-scale_2021}. Furthermore, the surrogate strategy presented in this section can be in principle expanded to learn other key metabolic fluxes, even if they are not involved in the differential equations of the hybrid models. This can help to achieve continued insight into metabolism, e.g., for process monitoring, without the need to solve an FBA problem at every sampling point. In this work, however, we focus only on obtaining machine-learning surrogates of FBA for $\bm{v_\mathrm{ext}}$.

\section{Optimization-based feedback control and estimation using the hybrid physics-informed model}
\label{sec:contol_est}
\begin{figure*}[htb!]
  \begin{center}
    \includegraphics[scale=0.60]{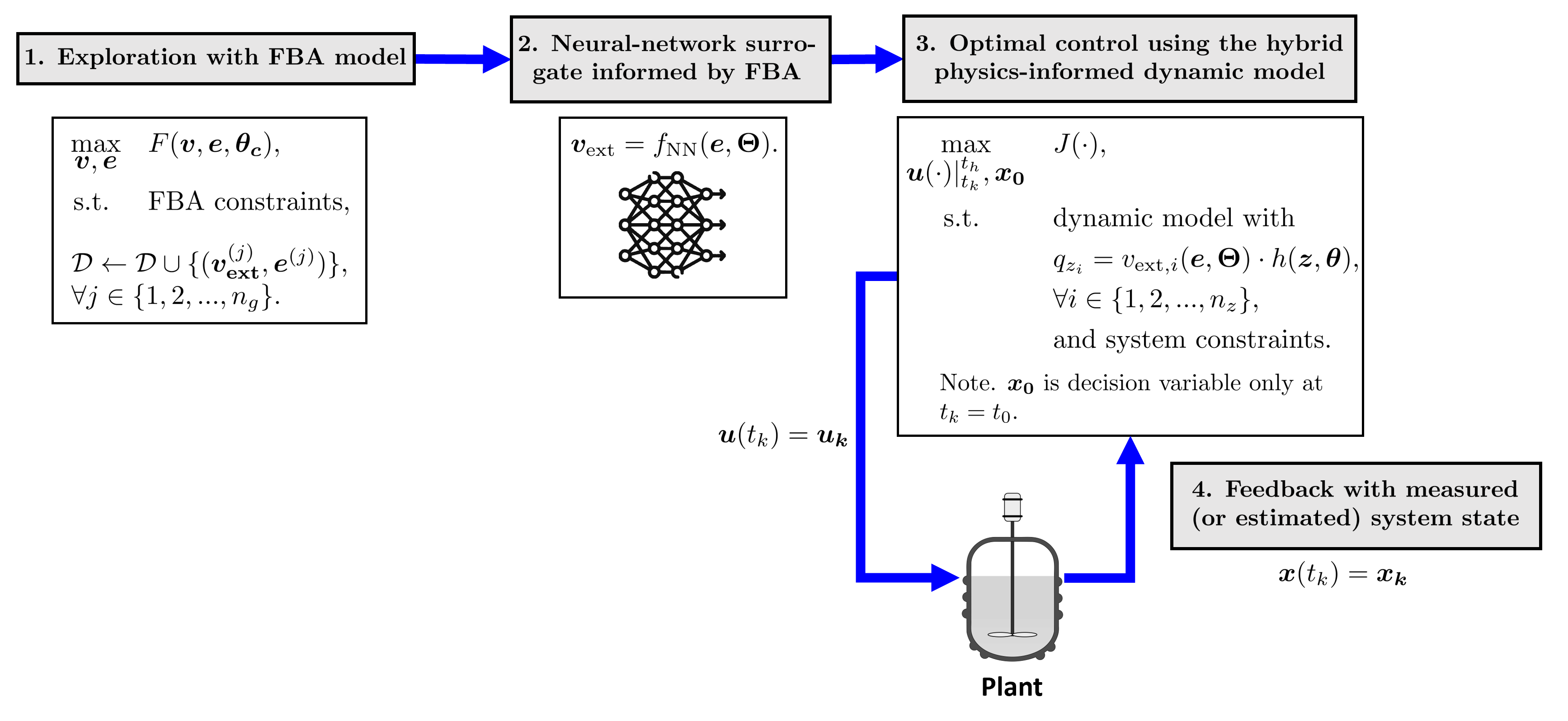}
    \caption{Flow diagram of our integrated hybrid physics-informed dynamic modeling and control strategy for metabolic cybergenetics. Details on steps 1 and 2 are provided in Section \ref{sec:hyb_pi_mod}, while details on steps 3 and 4 are provided in Section \ref{sec:contol_est}.}
    \label{fig:flowchart}
\end{center}
\end{figure*}

With the hybrid physics-informed dynamic model outlined in the previous section, we now aim to find the external input trajectories that maximize bioprocess production efficiency. To do so, we formulate a model-based optimal controller following a metabolic cybergenetic approach. The overall methodology is presented in Fig. \ref{fig:flowchart} and will be outlined in detail in this section.

\subsection{Model predictive control}
Let us assume for simplicity equidistant sampling points throughout the process. The sampling time is denoted as $t_k := k h_s$, $k \in \mathbb{N}_0$, with a sampling interval $h_s \in \mathbb{R}_{> 0}$. Also, $\bm{x} \in \mathbb{R}^{n_x}$ is the state vector, i.e., $\bm{x}:=[\bm{z}^\tran,\bm{e}^\tran]^\tran$ for prokaryotes and $\bm{x}:=[\bm{z}^\tran,\bm{e}^\tran,\bm{p}^\tran]^\tran$ for eukaryotes. Following a model predictive control approach \cite{rawlings_model_2020}, we repetitively solve an optimal control problem with the updated system state $\bm{x}(t_k)$, thereby incorporating state feedback. The state information can be obtained from hardware or soft sensors \cite{noll_history_2020}. The model predictive controller at time $t_k$ is formulated as the following optimization problem constrained by the hybrid physics-informed dynamic model:
\begin{align}
    {\max_{\bm{\mathfrak{u}}}} \quad & J(\cdot), \label{eq:MPC_obj} \\
    \text{s.t.} \quad & \text{Eqs. } \eqref{eq:z_dyn}-\eqref{eq:e_dyn}, \text{ if prokaryote}, \nonumber \\
    & \text{Eqs. } \eqref{eq:z_dyn_e}-\eqref{eq:p_dyn_e}, \text{ if eukaryote}, \nonumber \\
    & \text{Eqs. } \eqref{eq:qi_def}-\eqref{eq:ML_model_ext}, \nonumber \\
    & \bm{0} \leq \bm{g_p}(\bm{x},\bm{u},\bm{\theta},\bm{\Theta}), \label{eq:MPC_cons} \\
    & \bm{\mathfrak{u}} = \begin{cases} 
    \{\bm{u}(\cdot)|_{t_0}^{t_h}, \bm{x_0}\}, & \text{if } t_k=t_0, \\
    \bm{u}(\cdot)|_{t_k}^{t_h}, & \text{if } t_k>t_0, 
    \end{cases}\label{eq:MPC_inputs} \\
    & \bm{x}(t_k) = \bm{x_k}. \label{eq:MPC_xk}
\end{align}

The efficiency of the process is captured by the function $J(\cdot)$. In a production context, the optimal control problem can be formulated to maximize a metric such as volumetric productivity, economic profit, environmental impact, or a weighted combination of those. One could also think of minimizing the error in set-point- or trajectory-following-type problems, in case an optimal set-point or trajectory is \textit{a priori} known. An important consideration is made regarding the decision variables of the controller $\bm{\mathfrak{u}}$ (cf. Eq. \eqref{eq:MPC_inputs}). The dynamic decision variable of the optimization is the input function over $t \in [t_k,t_h]$, where $t_h$ is the time at the end of the prediction horizon. In batch setups, we often set $t_h=t_f$, where $t_f$ is the final batch time, hence a shrinking prediction horizon. In other circumstances, such as in continuous processes, one may consider that $t_h$ moves alongside the prediction horizon. 

Within the optimization, at $t_k=t_0$, one could also determine the optimal initial state of the system $\bm{x_0}$ (cf. Eq. \eqref{eq:MPC_inputs}). For example, one could start the process at specific concentrations of biomass, substrate, and manipulatable enzymes (via previous induction). However, remark that once the process has started, i.e., at $t_k > t_0$, we are only allowed to adjust the dynamic input. Possible (nonlinear) state or input constraints are also considered in the optimization problem. These are represented by the function $\bm{g_p}: \mathbb{R}^{n_x} \times \mathbb{R}^{n_u} \times \mathbb{R}^{n_{\theta}} \times \mathbb{R}^{n_{\Theta}} \rightarrow \mathbb{R}^{n_{g_p}}$ and can include economic, technical, environmental, and safety constraints.  It is also worth mentioning that if the optimization in Eqs. \eqref{eq:MPC_obj}-\eqref{eq:MPC_xk} is only solved at $t_k = t_0$ and does not incorporate state feedback, then it is regarded as an \textit{open-loop optimization}.

\subsection{Estimation of mRNA transcripts and manipulatable enzymes}
One of the challenges of (metabolic) cybergenetics is how to measure changes in intracellular states in real time for updating the model predictive controller. For example, accurately measuring concentrations of mRNA transcripts and intracellular enzymes often relies on offline technologies such as transcriptomics \cite{lowe_transcriptomics_2017} and proteomics \cite{shuken_introduction_2023}. The latter technologies are time-consuming and thus impractical to be employed for \textit{real-time} monitoring. One possibility for real-time monitoring of intracellular enzymes is to use fluorescent protein tags as reporters or biosensors  \cite{su_fluorescent_2005,lee_improved_2013}. However, the read-out can be challenging if the spectra of multiple reporters overlap. Additionally, the biosensor machinery might impose a resource burden on the cells. Also, biosensors are not without delays in their activation and inactivation. Consequently, in scenarios where measuring the intracellular states is challenging or impractical, we propose to estimate them via soft sensors.

Specifically, we propose an estimator based on moving horizon estimation and full information estimation that considers the dynamic model of the system and a history of past measurements, to infer the current state of the system. Our soft sensor at time $t_k$ is formulated in a discrete manner as follows \cite{rawlings_model_2020,pohlodek_flexible_2022}:
\begin{align}
    {\min_{\bm{x_{k-N}}, \bm{w}}} \quad & \left\Vert \bm{x_{k-N}}  - \bm{\bar{x}_{k-N}} \right\Vert_\mathbf{P}^2 + \sum_{i=k-N}^{k} \Vert \bm{\hat{y}_i} - \bm{y_i} \Vert_\mathbf{R}^2 + \sum_{i=k-N}^{k} \Vert \bm{w_i} \Vert_\mathbf{Q}^2, \label{eq:MHE_obj} \\
    \text{s.t.} \quad & \bm{x}_{i + h_s}= \bm{x_i} + \int_{t_i}^{t_i+h_s}\bm{f_d}(\bm{x},\bm{\hat{u}},\bm{\theta},\bm{\Theta}) \diff t  + \bm{w_i},\label{eq:x_dynamics_mhe} \\
    & \bm{y_i} = \bm{\eta}(\bm{x_i},\bm{\hat{u}_i},\bm{\theta},\bm{\Theta}), \label{eq:y_cons_mhe}\\
    & \bm{0} \leq \bm{g_p}(\bm{x_i},\bm{\hat{u}_i},\bm{\theta},\bm{\Theta}), \label{eq:g_cons_mhe} \\
    & \forall i \in \{k-N,...,k\}.\notag
\end{align}

Here $\bm{f_d}:\mathbb{R}^{n_x} \times \mathbb{R}^{n_u} \times \mathbb{R}^{n_\theta} \times \mathbb{R}^{n_\Theta} \rightarrow \mathbb{R}^{n_{x}}$ collects all the dynamic equations of the hybrid physics-informed dynamic model. $\bm{y} \in \mathbb{R}^{n_y}$ is the measurement vector and $\bm{\eta}:\mathbb{R}^{n_x} \times \mathbb{R}^{n_u} \times \mathbb{R}^{n_{\theta}} \times \mathbb{R}^{n_\Theta}  \rightarrow \mathbb{R}^{n_y}$ is the measurement function. $N \in \mathbb{N}$ is the length of the estimation window. $\bm{\hat{u}_i}$ and $\bm{\hat{y}_i}$ are the input and output measurements at time $t_i$, respectively. $\bm{{w}_i} \in \mathbb{R}^{n_x}$ is the state noise vector at time $t_i$, thus $\bm{w}:=\{\bm{w_i}, \, \forall i \in \{k-N,...,k\}\}$. 

Furthermore, $\bm{\bar{x}_{k-N}}$ is the \textit{best} guess of the state at the beginning of the estimation window. $\mathbf{P}$, $\mathbf{R}$ and $\mathbf{Q}$ are weighting matrices of appropriate dimensions. $\Vert \bm{a} \Vert_\mathbf{A}^2 := \bm{a}^\tran \mathbf{A} \bm{a}$ is the weighted squared norm of a vector $\bm{a}$ with respect to a weight matrix $\mathbf{A}$. The first term of the objective function in \eqref{eq:MHE_obj} represents the \textit{arrival cost} at the beginning of the estimation window. This serves as a kind of "memory" of past points left behind by the estimation window. The second term of the objective function represents the \textit{measurement error}, while the third term represents the \textit{process noise} or \textit{state disturbances}. The decision variable of the estimator is the initial state at the beginning of the estimation window $\bm{x_{k-N}}$ and the state noise $\bm{w}$, which can be used to regenerate the \textit{filtered}/\textit{smoothed} system dynamics via Eq. \eqref{eq:x_dynamics_mhe}. Note that even if the states can be measured through biosensors or other means, the estimator can still be beneficial to \textit{filter} possible noisy measurements.

Note that uncertain parameters can in principle also be estimated with the described soft sensor. However, for simplicity, here we assume constant model parameters in the estimation. It is also worth noting that the estimation window \textit{grows} in time, i.e., the number of data points within the estimation window increases, until the length $N$ is reached. Afterward, the estimation window only \textit{moves}, and the estimator is regarded as a \textit{moving horizon estimator}. If we choose $N$ such that the estimation window always starts from $t_0$, the estimation window always grows, and we refer to this estimator as a \textit{full information estimator}. We employ a full information estimator in the case study presented in the next section. Although here we consider equidistant sampling points, multi-rate estimators are also possible \cite{elsheikh_comparative_2021}.

\section{Results and discussion}
\label{sec:case_study}
\subsection{Computational case study: itaconate biosynthesis by \textit{E. coli} with optogenetic control}
\label{sec:itaconate_case_study}
As a starting point for our metabolic cybergenetic case study, we consider \textit{E. coli} ita36A, a strain engineered for itaconate biosynthesis from glucose under aerobic conditions \cite{harder_temperaturedependent_2018}. \textit{E. coli} ita36A features deletions of the genes \textit{aceA} (isocitrate lyase), \textit{pta} (phosphate acetyltransferase), \textit{pykF} (pyruvate kinase I), and \textit{pykA} (pyruvate kinase II). Additionally, the original strain harbors a plasmid that encodes the \textit{cadA} gene (cis-aconitate decarboxylase) from \textit{Aspergillus terreus}. Cis-aconitate decarboxylase enables the synthesis of itaconate from cis-aconitate, an intermediate metabolite of the tricarboxylic acid (TCA) cycle, through a decarboxylation reaction. The plasmid also encodes gene \textit{gltA} (citrate synthase) from \textit{Corynebacterium glutamicum}, enhancing the flux through the upper branch of the TCA cycle. It is important to note that both \textit{cadA} and \textit{gltA} are constitutively expressed in \textit{E. coli} ita36A. 

Furthermore, the native \textit{icd} (isocitrate dehydrogenase) in \textit{E. coli} ita36A has been replaced by a synthetic genetic switch that regulates \textit{icd} transcription via temperature control. The enzyme isocitrate dehydrogenase modulates the flux of the TCA cycle downstream isocitrate. Therefore, as demonstrated by \citeauthor{harder_temperaturedependent_2018}, the temperature-regulation of the expression of \textit{icd} can enhance the volumetric productivity of itaconate in two-stage fermentation processes \cite{harder_temperaturedependent_2018}. This can be done by performing a growth phase (with \textit{icd} expression off at 37 \textdegree C) followed by a production phase (with \textit{icd} expression on at 28 \textdegree C). The flux through the TCA cycle is intrinsically linked to growth, supplying precursor metabolites, energy cofactors, and reducing equivalents essential for biomass synthesis. Thus, reducing the flux through the TCA cycle by suppressing \textit{icd} expression \textit{indirectly} increases the flux through itaconate biosynthesis, although at the expense of growth due to the less active TCA cycle. The temperature-inducible control of \textit{icd} showed promising results for enhancing the process volumetric productivity; however, fine-tuning the bioreactor's temperature in large-scale setups poses challenges, particularly due to heat-transfer limitations. Additionally, temperature is a non-orthogonal control input, potentially affecting the transcription of untargeted genes in \textit{E. coli} \cite{gadgil_transcriptional_2008}.

In our computational case study, we introduce several modifications to the original \textit{E. coli} ita36A strain \cite{harder_temperaturedependent_2018}. First, we consider the dynamic regulation of \textit{cadA} as an alternative and more \textit{direct} mechanism to balance flux between the TCA cycle (\textit{growth mode}) and itaconate biosynthesis (\textit{production mode}). An increase in the flux of cis-aconitate decarboxylase is expected to divert carbon flux from the TCA cycle toward itaconate biosynthesis, as predicted by FBA simulations (cf. Section \ref{sec:cis-aconitate-decarboxylase-flux}). In this modified strain, \textit{icd} is constitutively expressed. For experimental implementations, however, it is crucial to evaluate whether the cell will indeed reduce the flux through the TCA cycle in favor of itaconate synthesis with increased expression of \textit{cadA}. There is a well-grounded possibility that the evolutionary driving force for cellular growth will make the TCA cycle preferable, despite increased expression of \textit{cadA}. \textit{If such a phenotype is observed}, one could further consider simultaneously inducing \textit{cadA} externally and repressing \textit{icd} expression to achieve tighter control. Having said that, \textit{for simplicity of demonstration of the hybrid physics-informed modeling and control cybergenetic framework}, we assume that regulating \textit{cadA} alone is sufficient to balance the flux through the TCA cycle and itaconate biosynthesis.

\begin{figure}[htb!]
  \begin{center}
    \includegraphics[scale=0.5]{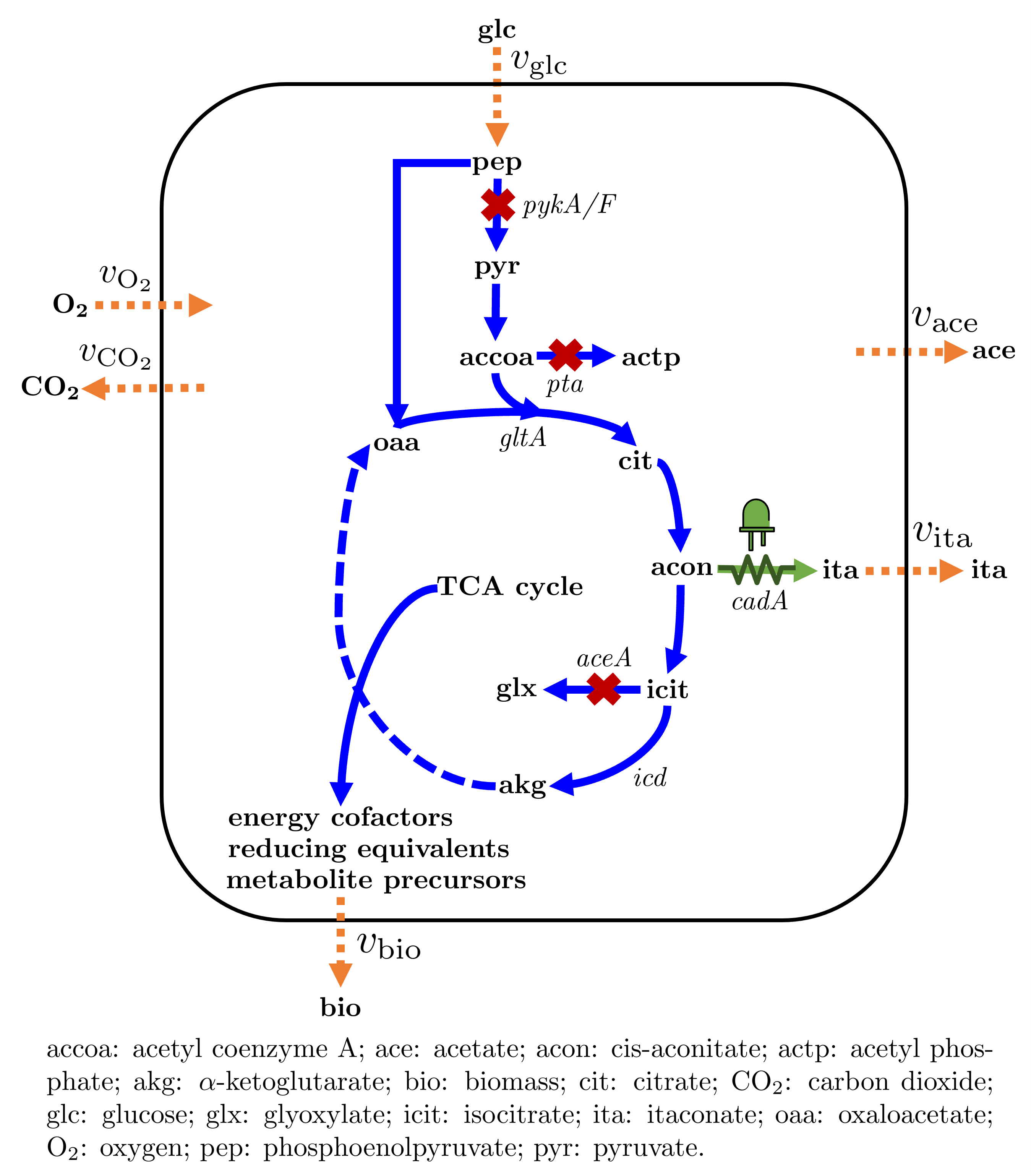}
    \caption{Scheme of \textit{E. coli} vOpt for itaconate biosynthesis. We represent gene deletions with red crosses. We indicate in italics genes encoding relevant enzymes. Orange dotted arrows denote metabolic exchange reactions. The green arrow with a resistor shape represents the reaction associated with the light-inducible cis-aconitate decarboxylase. $v_i \in \mathbb{R}$ denotes the metabolic exchange flux of the external species $i$.}
    \label{fig:pathway}
\end{center}
\end{figure}

Secondly, this computational study incorporates an optogenetic gene expression system to regulate the expression of \textit{cadA}. Utilizing light as a gene expression control input offers enhanced tunability and orthogonality compared to other inducers such as temperature. A simplified diagram illustrating the metabolism of this \textit{E. coli} strain for itaconate biosynthesis is presented in Fig. \ref{fig:pathway}. Throughout this study, our adapted \textit{E. coli} strain with optogenetic regulation is referred to as \textit{E. coli} vOpt, in which \textit{cadA} expression is activated by light in the fermentation.

\subsection{Assumptions and remarks on the hybrid physics-informed model}
Since \textit{E. coli} vOpt is a prokaryotic organism, the hybrid physics-informed metabolic model, as described by Eqs. \eqref{eq:z_dyn}-\eqref{eq:x0}, is applicable to our case study. We consider four external states: the concentrations of glucose, itaconate, acetate, and biomass; hence, the state vector is denoted as $\bm{z} := [z_{\mathrm{glc}}, z_{\mathrm{ita}}, z_{\mathrm{ace}}, b]^\top$. For simplicity, gas exchange is not included in the model and is assumed not to be a rate-limiting factor. Enzyme \textit{cadA} is the only enzyme under external control of gene expression, thus $\bm{e} := e_{\mathrm{cadA}}$, $\bm{q_{e_p}}:=q_{e_\mathrm{cadA}}$, and $\bm{d_{e}}:=d_{e_\mathrm{cadA}}$. The input-dependent expression rate of \textit{cadA}, $q_{e_{\mathrm{cadA}}}$, is modeled using the Hill activation function for lumped transcription and translation:
\begin{equation} \label{eq:hill_activation}
\begin{aligned}
q_{e_\mathrm{cadA}}=\theta_1+\theta_2 \frac{u_\mathrm{cadA}^{\theta_3}}{\theta_4^{\theta_3}+u_\mathrm{cadA}^{\theta_3}},
\end{aligned}
\end{equation}
with parameters $\theta_3 = 2.0780\, \mathrm{[dimensionless]}$ and $\theta_4 = 0.3799\, \mathrm{W/m^2}$ based on the values reported for the CcaS-CcaR 2.0 two-component optogenetic system  \cite{schmidl_refactoring_2014}. Here the input $u_{\mathrm{cadA}} \in \mathbb{R}_{\geq 0}$ is the intensity of green light ($\lambda_\mathrm{max}=520 \, \mathrm{nm}$) in $\mathrm{W/m^2}$, triggering \textit{cadA} transcription, hence $\bm{u}:=u_\mathrm{cadA}$. See Ref. \citenum{schmidl_refactoring_2014} for additional details on the mechanism of the CcaS-CcaR 2.0 optogenetic system. For simplicity, we assume no constitutive expression of the target enzyme, setting $\theta_1 = 0$. The parameter $\theta_2 \in \mathbb{R}_{\geq 0}$ is the maximal input-dependent rate of expression in $\mathrm{mmol/g_b/h}$. In later sections, we will compare different values of $\theta_2$, which represent various \textit{design} parameter values of the optogenetic gene expression system. In this work, we assume homogeneous light penetration inside the bioreactor, however, non-homogeneous light penetration can be also considered as in Ref. \citenum{espinel_cybergenetic_2023}.

The degradation rate of enzyme \textit{cadA} is modeled using first-order kinetics. Therefore:
\begin{equation} \label{eq:e_deg_cs}
\begin{aligned}
d_{e_\mathrm{cadA}}=\theta_5,
\end{aligned}
\end{equation}
where $\theta_5 = 0.6931\, \mathrm{h^{-1}}$ assuming an average protein half-life time of 1 $\mathrm{h}$ in \textit{E. coli} \cite{milo_cell_2016}.

The function $h$ in Eq. \eqref{eq:qi_def} considers limitation by the substrate glucose,  modeled using a hyperbolic (Monod) function, and inhibition by the by-product acetate, modeled using a hyperbolic decay function. Therefore, 
\begin{equation} \label{eq:h_cs}
\begin{aligned}
h = \left( \frac{z_\mathrm{glc}}{\theta_6 + z_\mathrm{glc}}\right) \left(\frac{\theta_7}{\theta_7 + z_\mathrm{ace}}\right),
\end{aligned}
\end{equation}
where the parameter $\theta_6 = 2.964 \times 10^{-04} \, \mathrm{mmol/L}$ is the assumed substrate affinity constant in the fermentation of glucose by \textit{E. coli} from previous determinations \cite{senn_growth_1994}. The parameter $\theta_7 = 134.63 \, \mathrm{mmol/L}$ was fitted using SciPy's \textit{curve\_fit} module \cite{2020SciPy-NMeth} with experimental data of acetate inhibition on \textit{E. coli}'s growth at pH 7.4 \cite{pinhal_acetate_2019}.

To relate the intracellular concentration of the enzyme \textit{cadA} to the attainable metabolic flux through the itaconate biosynthesis reaction (cf. Eq. \eqref{eq:e_k_cat}), we consider a catalytic constant $k_\mathrm{cadA} = 66,240\, \mathrm{h^{-1}}$ reported for the \textit{Aspergillus} version of the enzyme \cite{chen_crystal_2019}.

\subsection{Neural-network surrogates informed by FBA}
Here, we explain how we built the physics-informed component of our hybrid dynamic model (cf. Eq. \eqref{eq:ML_model_ext}) for \textit{E. coli} vOpt. First, we need a suitable metabolic network such that we can run FBA simulations under different levels of manipulatable flux values as described in the optimization problem in Eqs. \eqref{eq:FBA_cost_k}-\eqref{eq:e_cons_k}. Thus, we employed a version of the metabolic network EColiCore2 containing the heterologous reaction catalyzed by \textit{cadA} provided in Ref. \citenum{harder_model-based_2016}. The metabolic network comprises 115 reactions and 93 metabolites. Note that the original EColiCore2 \cite{hadicke_ecolicore2_2017}, derived from the genome-scale model iJO1366 of \textit{E. coli} \cite{orth_comprehensive_2011}, is a reduced metabolic network representing the cell's central metabolism.

\subsubsection{Preliminary accuracy check of the constraint-based model}
To validate the accuracy of the metabolic network to be used in our case study with \textit{E. coli} vOpt, we first ran FBA simulations using COBRApy \cite{ebrahim_cobrapy_2013} constrained by experimentally-reported metabolic exchange fluxes for \textit{E. coli} ita36A with temperature-dependent dynamic control of the TCA cycle \cite{harder_temperaturedependent_2018}. The objective function of the FBA was set to maximize the biomass metabolic flux $v_{\mathrm{bio}}$. As shown in Table \ref{tbl:FBA_validation}, the biomass metabolic fluxes predicted by FBA closely matched the experimentally-reported values during both growth and production phases by \textit{E. coli} ita36A. Therefore, we deemed this version of EColiCore2 suitable for testing our hybrid metabolic cybergenetic framework with \textit{E. coli} vOpt. Note that only constraints represented by Eqs. \eqref{eq:SV_cons_k} and \eqref{eq:V_cons_k} were considered in the preliminary accuracy check of the FBA model outlined in Table \ref{tbl:FBA_validation}, i.e., no enzyme-related constraints were taken into account.

\begin{table}[htb!]
  \centering
    \caption{Preliminary accuracy check of the proposed constraint-based model to be used in our case study with \textit{E. coli} vOpt. FBA simulations were performed considering \textit{E. coli} ita36A, with temperature-dependent dynamic control of the TCA cycle, constrained by experimentally-measured metabolic exchange fluxes. Exp.: experimental; pred.: predicted.} 
    \label{tbl:FBA_validation}
  \begin{threeparttable}
    \begin{tabular}{llll}
      \hline
      Exp. fluxes\tnote{a}\, [$\mathrm{mmol/g_b/h}$] & Exp. $v_{\mathrm{bio}}$\tnote{a}\, [$\mathrm{h^{-1}}$] & Pred. $v_{\mathrm{bio}}$\tnote{b}\, [$\mathrm{h^{-1}}$] & Reported phase\tnote{a}\\
      \hline
      $v_{\mathrm{glc}}=3.48$, $v_{\mathrm{ita}}=0.17$   & 0.25 $\pm$ 0.01 & 0.27 & Growth\\
      $v_{\mathrm{glc}}=0.85$, $v_{\mathrm{ita}}=0.43$   & 0.03 $\pm$ 0.00 & 0.01 & Production\\
      \hline
    \end{tabular}
    \begin{tablenotes}
\item[a]\footnotesize{Based on the experimental data for \textit{E. coli} ita36A provided in Ref. \citenum{harder_temperaturedependent_2018}.} 
\item[b]\footnotesize{For these FBA simulations, we assumed an ATP maintenance demand of 8.4 $\mathrm{mmol/g_b/h}$ and a formate excretion rate equal to zero based on the phenotype of previous iterations of \textit{E. coli} ita36A \cite{harder_model-based_2016}. We also set the exchange flux of glutamate, a possible way to replenish $\alpha$-ketoglutarate in the TCA, equal to zero. Gene knock-outs were simulated by setting to zero the metabolic fluxes associated with the genes \textit{aceA}, \textit{pta}, \textit{pykF}, and \textit{pykA}.}
    \end{tablenotes}
  \end{threeparttable}
\end{table}

\subsubsection{Systematic exploration of cis-aconitate decarboxylase flux manipulation}
\label{sec:cis-aconitate-decarboxylase-flux}
Prior to systematically exploring manipulatable intracellular flux values in \textit{E. coli} vOpt, we constrained the FBA model to match the expected phenotype. We constrained the FBA model for \textit{E. coli} vOpt with an ATP demand of 8.4 $\mathrm{mmol/g_b/h}$ and a formate exchange flux equal to zero, following a similar rationale as in the FBA simulations with \textit{E. coli} ita36A in Table \ref{tbl:FBA_validation}. The \textit{upper bound} of the glucose exchange flux was set to 3.48 $\mathrm{mmol/g_b/h}$, matching the experimental growth scenario in Table \ref{tbl:FBA_validation}, while the fluxes associated with genes knock-outs (\textit{aceA}, \textit{pta}, \textit{pykF}, and \textit{pykA}) were set equal to zero.

We consider only one manipulatable flux $v_\mathrm{cadA}$, namely the flux catalyzed by enzyme \textit{cadA}. The set of explored manipulatable flux values for this enzyme is denoted as $\mathbb{V}_\mathrm{man,cadA}$, hence $\mathbb{G}=\mathbb{V}_\mathrm{man,cadA}$ (cf. Eq. \eqref{eq:cartesian_product}). The set $\mathbb{V}_\mathrm{man,cadA}$ was defined as $\mathbb{V}_\mathrm{man,cadA} := \{0.000, 0.007, 0.014, \ldots, 3.476
\}$, covering the \textit{feasible} space of growth-production yield trade-offs. With $\mathbb{G}$ defined, we systematically solved the FBA problem, as detailed in the optimization problem in Eqs. \eqref{eq:FBA_cost_k}-\eqref{eq:e_cons_k}, to fill our dataset represented by Eq. \eqref{eq:dataset}.

\begin{figure}[htb!]
    \begin{center}    
        \subfigure[Relevant metabolic exchange fluxes]{\includegraphics[scale=0.7]{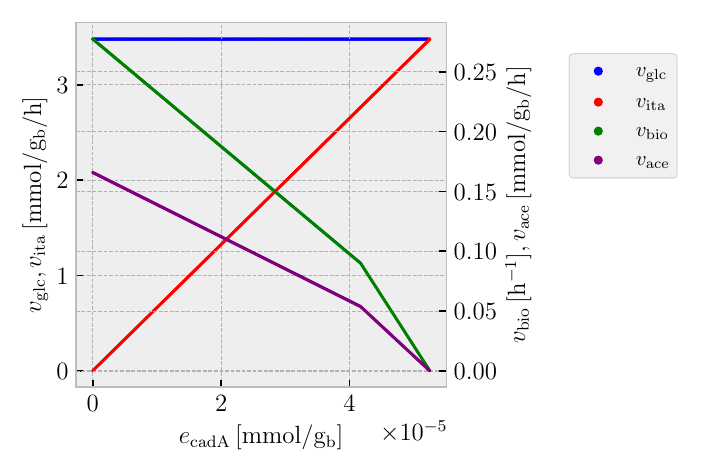}}
        \subfigure[Yield trade-off between growth and production]{\includegraphics[scale=0.7]{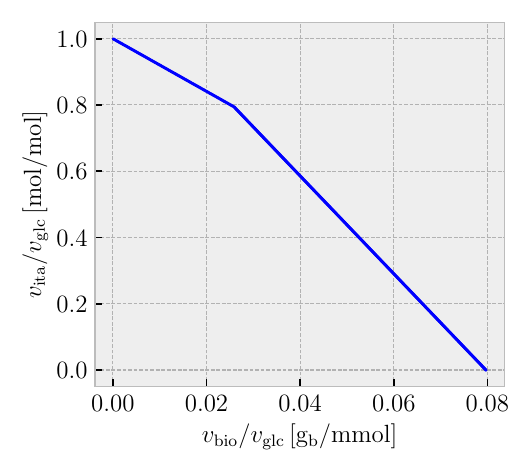}} 
        \caption{(a) Changes of metabolic exchange fluxes with varying intracellular cis-aconitate decarboxylase concentration predicted by the FBA model. (b) Solution space of the yield trade-off between itaconate and biomass synthesis from the exploration with FBA. glc: glucose, ita: itaconate, bio: biomass, ace: acetate, cadA: cis-aconitate decarboxylase.}
        \label{fig:fluxes}
    \end{center}
\end{figure}

Upon systematic exploration, Figure \ref{fig:fluxes}-a illustrates the predicted impact of intracellular concentration of enzyme \textit{cadA} on the metabolic exchange fluxes predicted by the FBA model. It is observed that the exchange flux of itaconate increases linearly with the concentration of enzyme \textit{cadA}, aligning with the assumed enzyme-capacity relationship described by Eq. \eqref{eq:e_k_cat}. As expected, this comes at the expense of a reduced biomass flux. Similarly, a decrease in the acetate exchange flux is observed with increasing concentrations of enzyme \textit{cadA}, despite the fact that the acetate exchange flux is already \textit{minimal} due to the \textit{pta} knock-out (the residual acetate in the FBA model originates from the synthesis reactions of the amino acids arginine and cysteine). In all scenarios, the FBA model predicts a glucose exchange flux of 3.48 $\mathrm{mmol/g_b/h}$, consistent with the flux's upper limit. The explored metabolic trade-off between growth and production is presented in Fig. \ref{fig:fluxes}-b. It becomes clear that the defined minimum and maximum values in $\mathbb{V}_\mathrm{man,cadA}$ were chosen to cover the range from the maximum biomass flux ($v_\mathrm{bio} = 0.277\, \mathrm{h^{-1}}$), with no itaconate exchange flux ($v_\mathrm{ita} = 0.000\, \mathrm{mmol/g_b/h}$ at $v_\mathrm{cadA}=0.000\, \mathrm{mmol/g_b/h}$), to the maximum itaconate exchange flux ($v_\mathrm{ita} = 3.476\, \mathrm{mmol/g_b/h}$), with no growth ($v_\mathrm{bio} = 0.000\, \mathrm{h^{-1}}$ at $v_\mathrm{cadA}=3.476\, \mathrm{mmol/g_b/h}$).

\begin{figure}[htb!]
    \begin{center}    
        \subfigure[Biomass]{\includegraphics[scale=0.5]{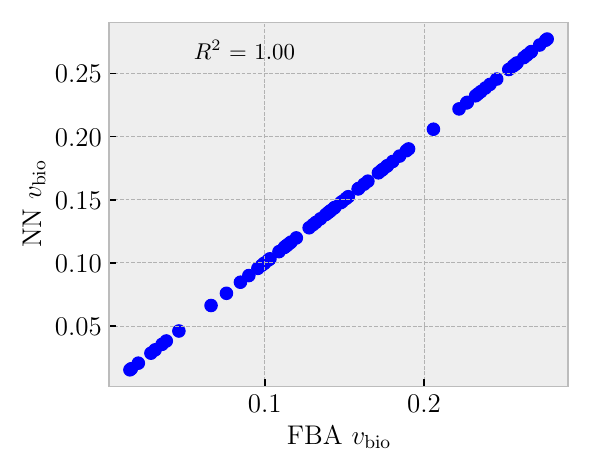}}
        \subfigure[Acetate]{\includegraphics[scale=0.5]{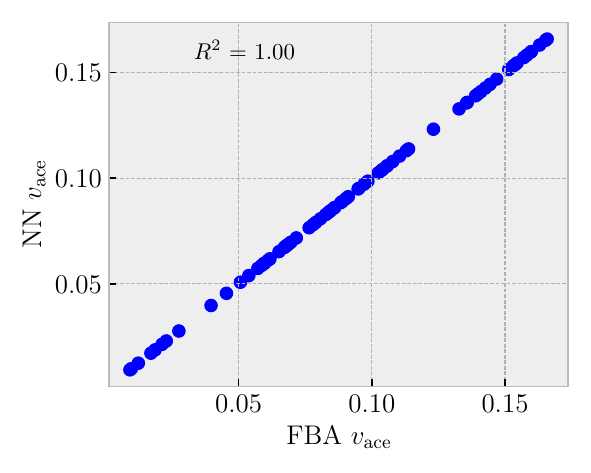}} 
        \subfigure[Itaconate]{\includegraphics[scale=0.5]{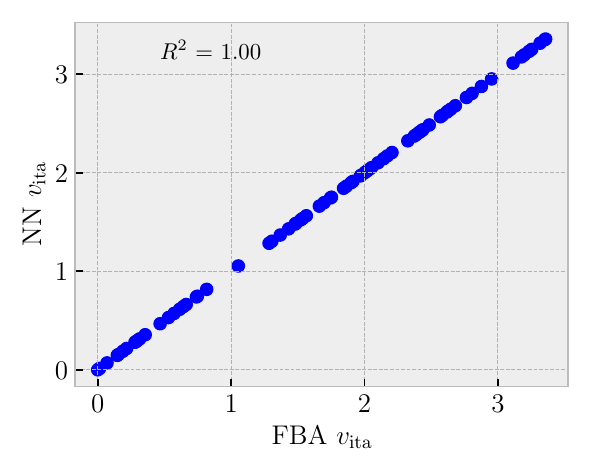}} 
        \caption{Parity plots of relevant metabolic exchange flux values for (a) biomass, (b) acetate, and (c) itaconate as predicted by FBA against the values as predicted by the neural network (NN).}
        \label{fig:parity_plots}
    \end{center}
\end{figure}

With the data generated during the systematic exploration, we trained a fully connected feedforward neural network using the MSE loss function represented by Eq. \eqref{eq:mse}. To do so, we used the Python toolbox HILO-MPC \cite{pohlodek_flexible_2022} interfaced with PyTorch \cite{paszke_pytorch_2019}. The goal was to generate Eq. \eqref{eq:ML_model_ext}, the physics-informed machine-learning part of the model. As the input feature, we considered the intracellular concentration of enzyme \textit{cadA}, and as output labels, the metabolic exchange fluxes of itaconate, acetate, and biomass. The metabolic exchange flux of glucose was not included as a label because it remained constant throughout the FBA exploration (see Fig. \ref{fig:fluxes}-a), and was therefore assumed constant in Eq. \eqref{eq:qi_def}. We shuffled the dataset, populated with the FBA solutions from the exploration phase, and reserved 15 \% of the data for testing the model's quality post-training. Of the remaining 85 \% of the data, 80 \% was used for training and 20 \% for validation, employing early stopping with a patience setting of 30.  To find the most suitable neural network hyperparameters, we performed a grid search considering different activation functions, number of layers, number of neurons per layer, and learning rates. Using the independent testing dataset to discriminate between trained neural networks, we selected an architecture of 2 hidden layers, 3 neurons per layer, and the ReLU (rectified linear unit) activation function, trained with a learning rate of 0.01. For this configuration, the parity plots comparing the independent dataset with the neural network's predictions yielded coefficients of determination $R^2$ of 1.00 (Fig. \ref{fig:parity_plots}). This indicates that the neural-network surrogate was able to effectively learn the physics of the FBA model for the metabolic exchange fluxes as a function of the manipulatable enzyme concentration without overfitting. 

Bear in mind that the case study uses an FBA model constrained only by a set of \textit{linear} equations. As such, the change of the metabolic exchange fluxes, with respect to the manipulatable enzyme concentration, was expected to follow relatively simple relationships as shown in Fig. \ref{fig:fluxes}, hence the relatively simple architecture of the neural-network surrogate. More complex relationships may, however, arise with more detailed FBA models, for example, considering constraints based on resource allocation, thermodynamics, and regulation\cite{beard_energy_2002,shlomi_genomescale_2007,goelzer_cell_2011}. An analogous methodology as presented here can be followed in those cases. In general, we recommend performing always hyperparameter optimization to find the most suitable architecture of the neural networks; this includes \textit{validation} and early stopping to avoid overfitting, as well as \textit{testing} to ensure good generalization. When in doubt about which configuration of the final neural network to select, metrics such as the Akaike information criterion \cite{parzen_information_1998} can become handy to achieve a good balance between model complexity and model accuracy \cite{zhao_design_2008}.

\subsection{Evaluating design parameters of the optogenetic gene expression system}
\begin{figure}[htb!]
  \begin{center}
    \includegraphics[scale=0.6]{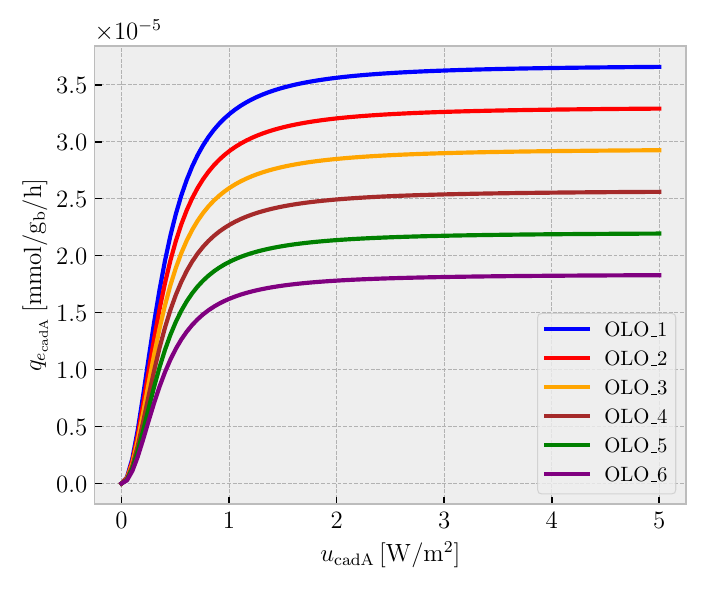}
    \caption{Hill activation function for the expression of the enzyme cis-aconitate decarboxylase across different design parameter values of $\theta_2$ in the considered optogenetic system (cf. Eq. \eqref{eq:hill_activation}). $\theta_2$ values; OLO\_1: $3.674\times 10^{-05}$, OLO\_2: $3.306\times 10^{-05}$, OLO\_3: $2.939\times 10^{-05}$, OLO\_4: $2.572\times 10^{-05}$, OLO\_5: $2.204\times 10^{-05}$, and OLO\_6: $1.837\times 10^{-05}$ $\mathrm{mmol/g_b/h}$.}
    \label{fig:par_design}
\end{center}
\end{figure}

With the neural-network surrogate now trained, our hybrid physics-informed dynamic metabolic model is almost complete. The model still has the parameter $\theta_2$ as a free design parameter of the optogenetic gene expression system (cf. Eq. \eqref{eq:hill_activation}). Therefore, to aid in the \textit{in silico} design of the gene expression system, we assessed the performance of various $\theta_2$ values in open-loop optimization problems. That is, the optimization problem in Eqs. \eqref{eq:MPC_obj}-\eqref{eq:MPC_xk} is solved just once at $t_0$, without state feedback, and assuming no mismatch between the plant's and controller's model. Specifically, we considered $\theta_2$ values of $3.674\times 10^{-05}$, $3.306\times 10^{-05}$, $2.939\times 10^{-05}$, $2.572\times 10^{-05}$, $2.204\times 10^{-05}$, and $1.837\times 10^{-05}$ $\mathrm{mmol/g_b/h}$. The latter values are within the range of orders of magnitude used in previous simulation studies of gene expression \cite{espinel_cybergenetic_2023}. The Hill activation function $q_{e_\mathrm{cadA}}$ for these different scenarios is plotted in Fig. \ref{fig:par_design}. The primary distinction among these scenarios is the maximum achievable expression rate, effectively the \textit{strength} of the gene expression system. The objective function was set to maximize itaconate concentration at the end of the batch, hence $J(\cdot)= z_{\mathrm{ita}}(t_f)$. We also set the upper bound of the input to $5 \, \mathrm{W/m^2}$ as the function $q_{e_\mathrm{cadA}}$ plateaus at this level for all values of $\theta_2$. Note that we only optimized the dynamic light input, while the initial conditions of the plant were \textit{a priori} given. We considered 24 equidistant inputs with an interval size of 1 h. Such discretization of the input function aids in approximating the otherwise infinite-dimensional function \cite{findeisen_introduction_2002}. The dynamic optimization using the hybrid physics-informed model was implemented in HILO-MPC  \cite{pohlodek_flexible_2022}.

\begin{figure}[htb!]
    \begin{center}    
        \subfigure[Light input]{\includegraphics[scale=0.6]{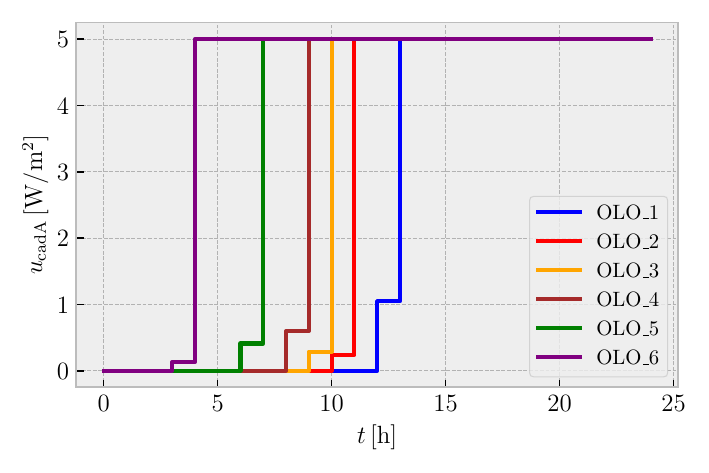}}
        \subfigure[Manipulatable enzyme]{\includegraphics[scale=0.6]{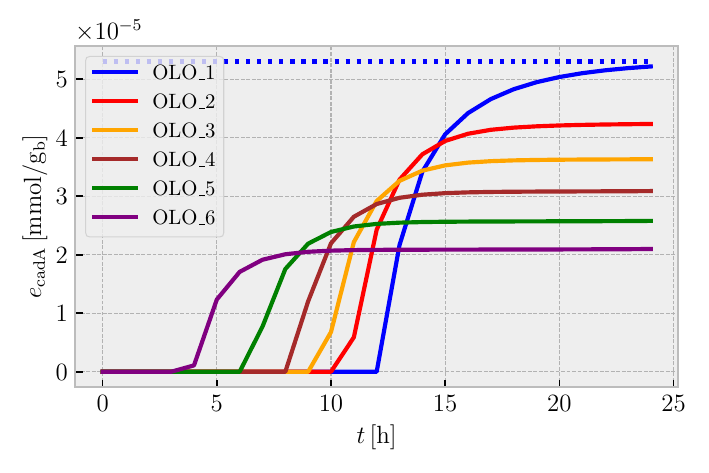}} \\  
        \vspace{-0.2cm}
        \subfigure[Biomass]{\includegraphics[scale=0.6]{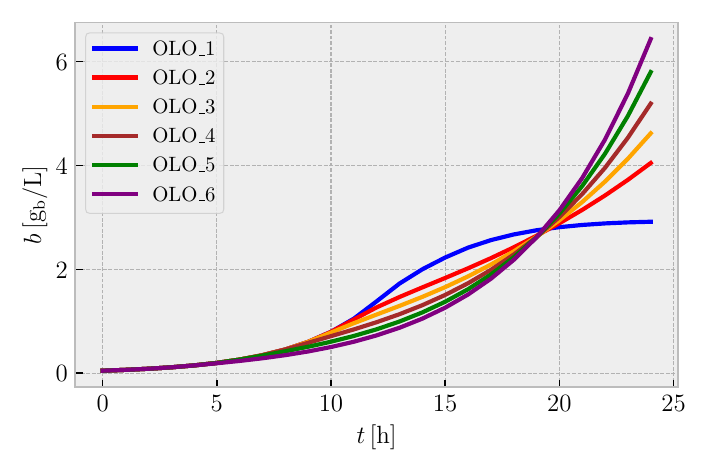}}
        \subfigure[Glucose]{\includegraphics[scale=0.6]{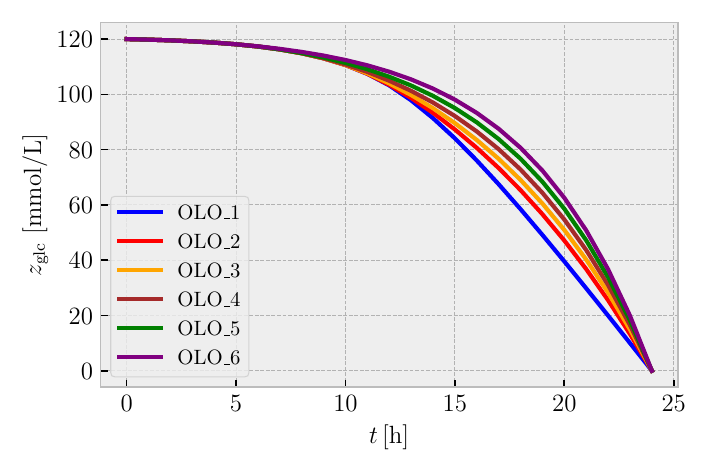}} \\       
        \vspace{-0.2cm}
        \subfigure[Acetate]{\includegraphics[scale=0.6]{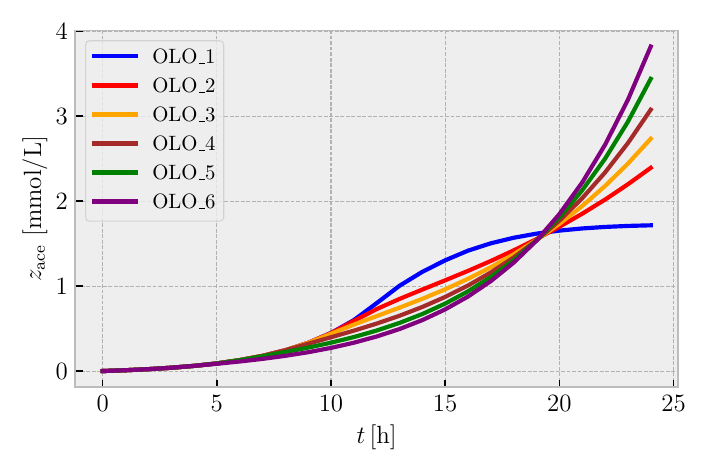}}
        \subfigure[Itaconate]{\includegraphics[scale=0.6]{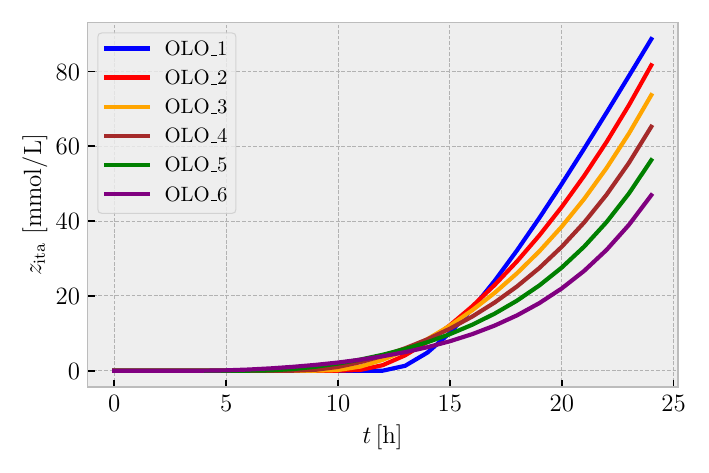}} \\  
        \caption{Open-loop optimizations using the hybrid physics-informed dynamic model with varying design parameter values of $\theta_2$ in the optogenetic system (cf. Eq. \eqref{eq:hill_activation}). We show the dynamic profiles for (a) the light input, (b) the manipulatable enzyme, (c) biomass, (d) glucose, (e) acetate, and (f) itaconate. It is assumed there is no mismatch between the controller's model and the plant's model. $\theta_2$ values; OLO\_1: $3.674\times 10^{-05}$, OLO\_2: $3.306\times 10^{-05}$, OLO\_3: $2.939\times 10^{-05}$, OLO\_4: $2.572\times 10^{-05}$, OLO\_5: $2.204\times 10^{-05}$, and OLO\_6: $1.837\times 10^{-05}$ $\mathrm{mmol/g_b/h}$. Initial conditions:  $z_\mathrm{glc}(t_0)=120 \, \mathrm{mmol/L}$, $z_\mathrm{ita}(t_0)=0.00 \, \mathrm{mmol/L}$, $z_\mathrm{ace}(t_0)=0.00 \, \mathrm{mmol/L}$, $b(t_0)=0.05 \, \mathrm{g_b/L}$, $e_\mathrm{cadA}(t_0)=0.00\times10^{-05} \, \mathrm{mmol/g_b}$. glc: glucose, ita: itaconate, bio: biomass, ace: acetate, cadA: cis-aconitate decarboxylase.}
        \label{fig:olo_dyn_scenarios}
    \end{center}
\end{figure}

The results of the open-loop optimizations are shown in Fig. \ref{fig:olo_dyn_scenarios}. In all scenarios, the light input follows a two-stage trajectory. Initially, the light input is off, completely repressing the expression of \textit{cadA}. Subsequently, the light input increases to its maximum allowable value. This aligns with an initial growth phase marked by maximal flux through the TCA cycle, thus no itaconate production. It is followed by a gradual increase in the intracellular concentration of enzyme \textit{cadA}, which enhances itaconate biosynthesis, at the expense of a reduced flux through the TCA cycle and biomass synthesis. The difference between input profiles is the triggering time for the second phase. The higher the value of $\theta_2$, the more delayed the triggering of the production phase. That is, with the enzyme being expressed at higher intracellular levels, an extended growth phase is \textit{affordable} with increasing $\theta_2$, allowing for significant biomass accumulation for enhanced volumetric productivity. Conversely, with lower values of $\theta_2$, less manipulatable enzyme accumulates intracellularly, necessitating a longer production phase to somewhat compensate for the reduced itaconate yields.

It is worth noting that the manipulatable enzyme concentration eventually plateaus, approaching a pseudo-steady-state intracellular level, influenced by the value of $q_{e_\mathrm{cadA}}$ (Fig. \ref{fig:par_design}). To prove this point, let us consider for simplicity a scenario where the manipulatable enzyme concentration is such that it leads to the maximum biologically-feasible flux of $v_\mathrm{cadA}$, hence $v_\mathrm{bio}=0$ and $\mu=0$. In this specific scenario, the pseudo-steady-state concentration of the enzyme $e_{\mathrm{cadA}_\mathrm{ss}}$ is:
\begin{equation}
\begin{aligned}
\label{eq:e_ss}
e_{\mathrm{cadA}_\mathrm{ss}} := \frac{q_{e_\mathrm{cadA}}}{d_{e_\mathrm{cadA}}}.
\end{aligned}
\end{equation}

Therefore, one can say that the higher the value of $q_{e_\mathrm{cadA}}$, the higher the intracellular pseudo-steady-state concentration of the enzyme. Note that the case with the maximum value of $\theta_2 = 3.674\times 10^{-05} \, \mathrm{mmol/g_b/h}$ closely aligns with the scenario described in the derivation of Eq. \eqref{eq:e_ss} at prolonged induction levels. To illustrate this, we have represented  $e_{\mathrm{cadA}_\mathrm{ss}}$ with a dotted blue line in Fig. \ref{fig:olo_dyn_scenarios}-b for the scenario with the highest value of $\theta_2$. There, the intracellular concentration of enzyme \textit{cadA} approaches the calculated value of $e_{\mathrm{cadA}_\mathrm{ss}}$ when the enzyme is fully induced.

In terms of biomass growth, only the scenario with the highest $\theta_2$ value exhibits a stagnation in biomass concentration towards the mid-to-end of the batch. This is explained by the fact that the enzyme \textit{cadA} accumulates to levels such that $v_\mathrm{bio} \approx 0$ (cf. Fig. \ref{fig:fluxes}-a). In the other scenarios, featuring decreasing values of $\theta_2$, the growth rate becomes correspondingly more exponential, a trend explained by a lower intracellular concentration of enzyme \textit{cadA} and a more active TCA cycle. Since the acetate metabolic flux is growth-coupled (see Fig. \ref{fig:fluxes}-a), the dynamic behavior of acetate mirrors that of biomass. In terms of itaconate, the product of interest, production only starts after the light input is switched on. As expected, the rate of itaconate production correlates with the intracellular concentration of enzyme \textit{cadA}. Therefore, higher itaconate titers are observed for the scenarios with higher enzyme \textit{cadA} accumulation. Naturally, enhanced titers are also enabled by the two-stage fermentation process, where the initial \textit{growth-dominant} stage facilitates sufficient biomass accumulation to support the subsequent \textit{production-dominant} phase.

Based on the predicted performance of the open-loop optimizations, we selected the highest value of $\theta_2 = 3.674\times 10^{-05} \, \mathrm{mmol/g_b/h}$ as the most suitable design parameter value for the optogenetic gene expression system. This value was chosen because it enables the coverage of the entire space of growth-production trade-offs, as presented in Fig. \ref{fig:fluxes}, a feature desirable for efficient dynamic metabolic control. We refer to the described hybrid physics-informed model, with the specified parameters, as the \textit{nominal model}. It should be noted that the results in this section do not account for system uncertainty, and therefore, the controller's model is assumed to precisely describe the plant's dynamics. That is, the nominal model applies to both the controller and the plant. In reality, bioprocesses are frequently subject to uncertainties, including model-plant mismatch and disturbances that need to be addressed with closed-loop control, which will be the focus of the next section. 

\subsection{Putting hybrid physics-informed metabolic cybergenetics into action}
We implemented a metabolic cybergenetic scheme to evaluate the performance of the bioprocess under system uncertainty. This scheme follows the optimization formulation presented in Eqs. \eqref{eq:MPC_obj}-\eqref{eq:MPC_xk}, which spans in the time domain from $t_k$ to $t_f$. Samples are taken hourly to re-optimize the system using the current state information at time $t_k$. As a source model-plant mismatch, we scaled the function $h$ in Eq. \eqref{eq:h_cs} by a factor of 1.04. This model is henceforth denoted as the \textit{modified model}.

\begin{figure}[htb!]
    \begin{center}    
        \subfigure[Light input]{\includegraphics[scale=0.6]{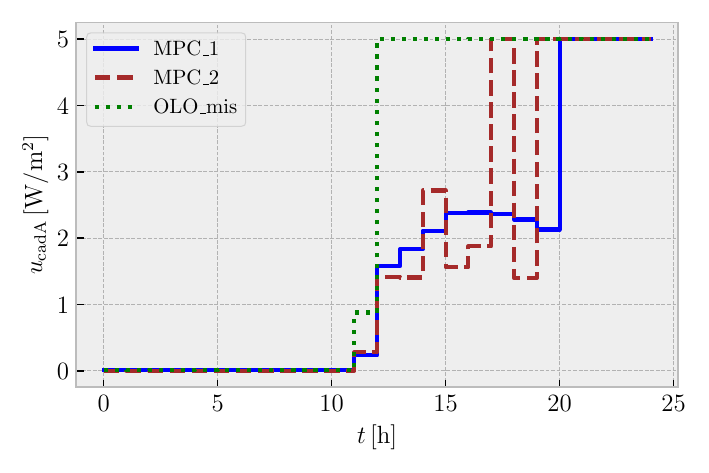}}
        \subfigure[Manipulatable enzyme]{\includegraphics[scale=0.6]{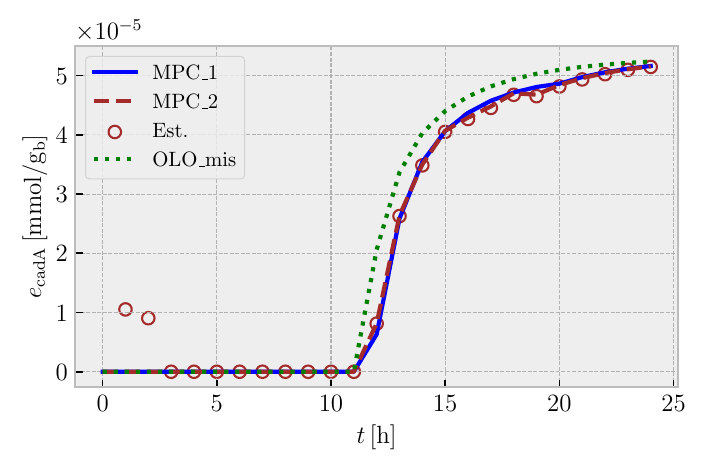}} \\  
        \vspace{-0.2cm}
        \subfigure[Biomass]{\includegraphics[scale=0.6]{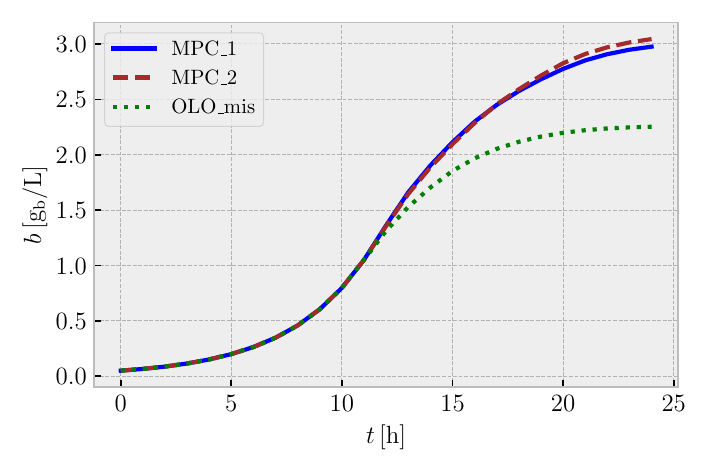}}
        \subfigure[Glucose]{\includegraphics[scale=0.6]{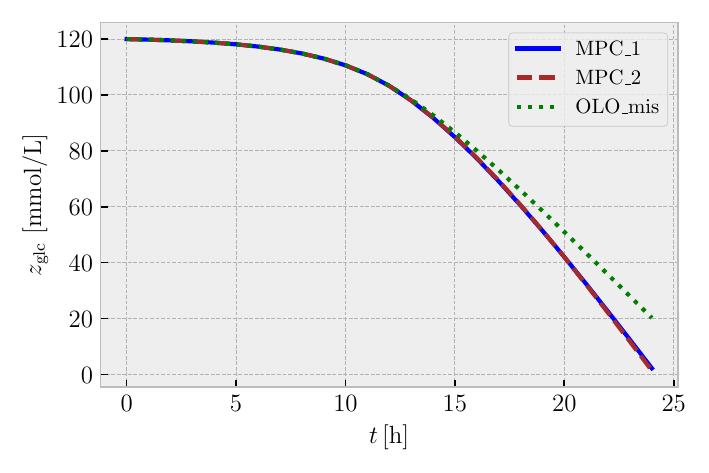}} \\       
        \vspace{-0.2cm}
        \subfigure[Acetate]{\includegraphics[scale=0.6]{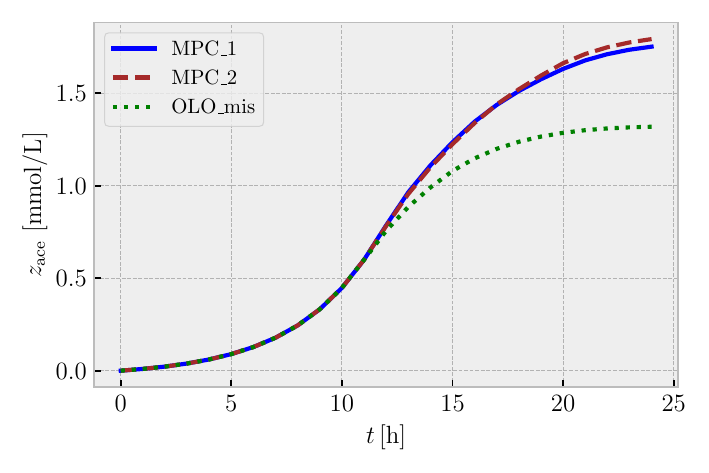}}
        \subfigure[Itaconate]{\includegraphics[scale=0.6]{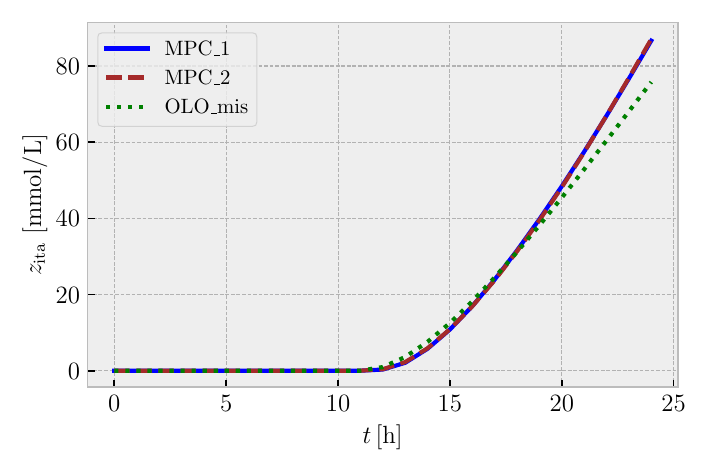}} \\  
        \caption{Model predictive control scenarios using the hybrid physics-informed dynamic model.  We show the dynamic profiles for (a) the light input, (b) the manipulatable enzyme, (c) biomass, (d) glucose, (e) acetate, and (f) itaconate. It is assumed that there is a mismatch between the controller's model and the plant's model. To ensure a fair comparison between scenarios, exact plant values are plotted. Initial conditions:  $z_\mathrm{glc}(t_0)=120 \, \mathrm{mmol/L}$, $z_\mathrm{ita}(t_0)=0.00 \, \mathrm{mmol/L}$, $z_\mathrm{ace}(t_0)=0.00 \, \mathrm{mmol/L}$, $b(t_0)=0.05 \, \mathrm{g_b/L}$, $e_\mathrm{cadA}(t_0)=0.00\times10^{-05} \, \mathrm{mmol/g_b}$.
       }
        \label{fig:mpc_dyn_scenarios}
    \end{center}
\end{figure}

In the first feedback control case, denoted as MPC\_1, we provided the controller with the modified model and used the nominal model to simulate the plant behavior. We also assumed full and accurate state information; that is, we considered all states measurable, including the intracellular enzyme \textit{cadA}, without noise at every sampling instance. Control simulation results for MPC\_1 are presented in Fig. \ref{fig:mpc_dyn_scenarios}. Note that we also plotted the open-loop control performance, denoted as OLO\_mis, for comparison; remark that this represents the case without state feedback. Although MPC\_1 is a very optimistic scenario, it provides a \textit{benchmark} for the performance of the model predictive controller. 

When analyzing MPC\_1 against OLO\_mis, the feedback controller effectively adapts the dynamic light input to compensate for the model-plant mismatch. The adjusted input facilitates greater biomass accumulation explained by a slightly lower intracellular concentration of enzyme \textit{cadA} after the onset of the induction. As a result, there is a higher volumetric productivity, resulting in a final itaconate titer of $86.64 \, \mathrm{mmol/L}$ in MPC\_1, which is $14.32 \, \%$ higher than the achieved with  OLO\_mis. Furthermore, MPC\_1 depletes up to $98.2\, \%$ of the initial glucose, compared to only $83.2 \, \%$ achieved with OLO\_mis.

In a second feedback control case, denoted as MPC\_2, we also provided the controller with the modified model, while we used the nominal model to simulate the plant behavior. However, unlike the external states, we considered that the intracellular concentration of enzyme \textit{cadA} could not be measured. Therefore, we used a full information estimator, as described in the optimization problem in Eqs. \eqref{eq:MHE_obj}-\eqref{eq:g_cons_mhe}, to reconstruct the system state.  We incorporated measurement noise by sampling the measurements from a Gaussian distribution with a 1.5 \% standard deviation with respect to the true plant value. We set $\mathbf{P}$ as a 5 $\times$ 5 diagonal matrix with diagonal elements equal to 10. Similarly, we set $\mathbf{R}$ as a 4 $\times$ 4 diagonal matrix with diagonal elements equal to 1,000. These weights were tuned to provide convergence and (to our judgment) good estimation quality. We omitted the term associated with state disturbances from the objective function of the state estimator as it did not make a difference in terms of the quality of the outcome. The \textit{measured} external states and the \textit{reconstructed} manipulatable enzyme were provided to the MPC controller at each sampling instance. Note that, like the controller, the estimator was provided with the modified model. The control simulation results for MPC\_2 are also plotted in Fig. \ref{fig:mpc_dyn_scenarios}.

When comparing MPC\_2 against OLO\_mis, the feedback controller is also able to compensate for the model-plant mismatch by adjusting the dynamic input. Like MPC\_1, MPC\_2 enables more biomass accumulation due to the slightly less intracellular concentration of enzyme \textit{cadA}  after the onset of the induction. In fact, the performance of MPC\_2 is comparable to MPC\_1, yielding a final itaconate concentration of $87.07 \, \mathrm{mmol/L}$, which is $14.88 \, \%$ higher than that achieved with OLO\_mis. Unlike OLO\_mis,  MPC\_2 is able to deplete up to $98.9\, \%$ of the initial glucose. Regarding the state estimator's performance, as shown in Fig. \ref{fig:mpc_dyn_scenarios}-b, the estimator accurately reconstructs the manipulatable enzyme concentration, closely matching the exact plant values. Only the first two estimates are slightly off, which can be explained by the few data points considered in those estimation instances.

Throughout this study, we have considered $\bm{\theta}$ as \textit{time-invariant} in the hybrid physics-informed model. However, doing so may neglect possible uncertainty arising from the explicitly modeled prior knowledge. Model uncertainty can also originate from the assumed values of the catalytic constants. With this in mind, the function $h$ in the control studies was scaled by 4 \%, which we deemed as a considerable model-plant mismatch judging from the difference in outcome between the open-loop and model predictive controller. The latter managed to address the introduced model-plant mismatch with an overall satisfactory performance. However, one should take into account that a higher model-plant mismatch may diminish the accuracy of the controller. These aspects, if not properly addressed, may affect the efficiency of the outlined model-based optimization and control strategy upon implementation.

To counteract the above-mentioned uncertainties, one option is to re-estimate $\bm{\theta}$ or a subset of it during the process operation inspired, e.g., by moving horizon estimation schemes \cite{jabarivelisdeh_adaptive_2020}. Sensitivity analysis for determining suitable parameters to be re-estimated may be appropriate. As such, this would enable \textit{online} adaptation of uncertain model parameters for increased accuracy of model-based optimization and control. Along the same line, data-driven methods can also be considered to tackle uncertainty in  $\bm{\theta}$ and in the hybrid physics-informed model in general. For example, $\bm{\theta}$ or a subset of it can be fully or partially treated as \textit{time-varying} and \textit{learned} from process data \cite{rogers_investigating_2023}. Similarly, the process data can be mined to learn the \textit{error} of the hybrid model to be then incorporated into the right-hand side of the differential equations \cite{morabito_efficient_2022,espinel-rios_batch--batch_2023,espinel-rios_experimentally_2024}. In this context, if one manages to capture the uncertainty of the hybrid model, e.g., via Gaussian processes, \textit{stochastic} optimization and control approaches could be unlocked, which offers in principle more robustness and can deal with probability constraints\cite{bradford_stochastic_2020,morabito_efficient_2022}.

\section{Conclusion}
This study outlines a hybrid physics-informed dynamic modeling framework for predictive control and estimation in metabolic cybergenetics. We embed the \textit{physics} of metabolic networks into the process rates of macro-kinetic models augmented with gene expression, which helps to account for intrinsic delays related to transcription and translation of manipulatable enzymes. Such models are built by employing machine-learning surrogates informed by FBA, wherein manipulatable intracellular enzymes serve as input features and relevant metabolic exchange fluxes as output labels. This approach ensures that the information from the system's metabolic network is not lost; rather, we extract valuable \textit{physics} from it to enhance structurally simpler dynamic models. In comparison to previous metabolic cybergenetic frameworks, particularly those utilizing dynamic constraint-based metabolic modeling \cite{espinel_cybergenetic_2023}, our hybrid machine-learning-supported modeling approach circumvents complex \textit{bilevel} optimization problems. That is, process control and estimation tasks can be formulated as \textit{single-level} optimizations, thereby simplifying the implementation of metabolic cybergenetics. Additionally, our framework reduces the number of states that need to be measured or estimated in real time during the operation of the bioprocess due to the smaller number of states involved in the models.

\begin{acknowledgement}
This work was supported by the U.S. Department of Energy, Office of Science, Office of Biological and Environmental Research (Award Number DE-SC0022155).
\end{acknowledgement}

\bibliography{cas-refs}

\end{document}